\documentclass[aps,prd,amsmath,superscriptaddress,%
               nobibnotes,nofootinbib,preprintnumbers,
               onecolumn,12pt,tightenlines]{revtex4}

\def\mysections#1{{\bf #1.} }

\usepackage{amssymb}
\usepackage{amsmath}
\usepackage{graphicx}
\usepackage{longtable}
\usepackage{verbatim}
\usepackage{amsfonts}
\usepackage[bookmarksdepth=2]{hyperref}

\newcommand{\be}{\begin{eqnarray}}
\newcommand{\ee}{\end{eqnarray}}
\newcommand{\bea}{\begin{eqnarray}}
\newcommand{\eea}{\end{eqnarray}}
\newcommand{\beq}{\begin{eqnarray}}
\newcommand{\eeq}{\end{eqnarray}}
\def\beqa{\begin{eqnarray}}
\def\eeqa{\end{eqnarray}}
\newcommand{\no}{\nonumber}
\def\lsim{\mathrel{\rlap{\lower4pt\hbox{\hskip1pt$\sim$}}
    \raise1pt\hbox{$<$}}}         
\def\gsim{\mathrel{\rlap{\lower4pt\hbox{\hskip1pt$\sim$}}
    \raise1pt\hbox{$>$}}}         

\def\R{\mathcal{R}}

\newcommand{\tce}{\frac{t_{\rm cool}(\R)}{t_{\rm esc}(\R)}}
\def\Xe{X_{\rm esc}}
\def\X{X_{\rm esc}}
\def\te{t_{\rm esc}}
\def\tc{t_{\rm cool}}
\def\nb{n_{\rm B}}
\def\nc{n_{\rm C}}

\def\rism{\rho_{\rm ISM}}
\def\x{(\R,\vec r,t)}
\def\xo{(\R,\vec r_\odot,t_\odot)}
\def\pb{\bar p}
\def\ep{e^+}
\def\ad{\rm\bar d}
\def\ah{\overline{\rm ^3He}}

\begin{document}

\vspace*{-30mm}

\title{\boldmath Cosmic-ray Antimatter
}

\author{Kfir Blum}\email{kfir.blum@weizmann.ac.il}
\affiliation{Dept. of Part. Phys. \& Astrophys., Weizmann Institute of Science, POB 26, Rehovot, Israel}
\affiliation{Theoretical Phys. Dept., CERN, Switzerland}
\author{Ryosuke Sato}\email{ryosuke.sato@weizmann.ac.il}
\affiliation{Dept. of Part. Phys. \& Astrophys., Weizmann Institute of Science, POB 26, Rehovot, Israel}
\author{Eli Waxman}\email{eli.waxman@weizmann.ac.il}
\affiliation{Dept. of Part. Phys. \& Astrophys., Weizmann Institute of Science, POB 26, Rehovot, Israel}

\vspace*{1cm}

\begin{abstract}
In recent years, space-born experiments have delivered new measurements of high energy cosmic-ray (CR) $\pb$ and $\ep$. In addition, unprecedented sensitivity to CR composite anti-nuclei $\ad$ and $\ah$ is expected to be achieved in the near future. We report on the theoretical interpretation of these measurements. While CR antimatter is a promising discovery tool for new physics or exotic astrophysical phenomena, an irreducible background arises from secondary production by primary CR collisions with interstellar matter. Understanding this irreducible background or constraining it from first principles is an interesting challenge. We review the attempt to obtain such understanding and apply it to CR $\pb,\,\ep,\,\ad,$ and $\ah$. 

Based on state of the art Galactic cosmic ray measurements, dominated currently by the AMS-02 experiment, we show that: (i) CR $\pb$ most likely come from CR-gas collisions; (ii) $\ep$ data is consistent with, and suggestive of the same secondary astrophysical production mechanism responsible for $\pb$ and dominated by proton-proton collisions. In addition, based on recent accelerator analyses we show that the  flux of secondary high energy $\ah$ may be observable with a few years exposure of  AMS-02. We highlight key open questions, as well as the role played by recent and upcoming space and accelerator data in clarifying the origins of CR antimatter.
\end{abstract}

\maketitle
\tableofcontents

\section{Introduction}\label{sec:0}
%

Cosmic ray (CR) physics is centred on the attempt to solve a set of basic open questions~\cite{Ginzburg:1976dj, Ginzburg:1990sk}. Where do CRs come from, namely, what are the sources of the bulk of the CR energy density? where are these sources located, and how do they accelerate particles? Narrowing to the Galactic energy range ($\sim$MeV to at least $\sim$PeV per particle): what role do CRs play in Galactic dynamics? How do CRs propagate and eventually escape from the Galaxy? Where and how does the transition to extra-galactic sources occur? 

From this wide perspective CR antimatter is an exotic phenomena, making an insignificant contribution to the total CR energy density (in the ballpark of 0.01 percent). 
Nevertheless, careful study of CR antimatter may reveal important insights to CR physics and beyond that, particle physics.  
Because considerable energy is required to create antimatter, it is a valuable probe of high energy processes. Because primordial antimatter is essentially absent before structure formation, its formation as secondary product in CR collisions with ambient gas makes the description of CR antimatter theoretically clean, decoupling the problem -- to some extent -- from core unknowns in the physics of CR sources and acceleration. 

As a result, one is led to a situation where although the sources of the bulk of the Galactic CRs -- primary protons, He, and other nuclei -- are essentially not understood, nevertheless the small residual accompanying radiation of $\pb$, essentially is. Using this theory control, spectral features or excess abundance of CR antimatter could lead to a first detection of exotic phenomena like dark matter annihilation or $e^\pm$ pair-production in pulsars. In addition, antimatter provides a unique handle on the physics of CR propagation.   

In this review we discuss CR antimatter. Aiming to jump directly to what we consider new and exciting developments, we leave out most of the basic CR physics background; as a partial list of useful books we recommend~\cite{Ginzburg:1990sk,Longair:1992ze,2005ppfa.book.....K,gaisserbook,Schlickeiser:2002pg}. Review papers are referred to where relevant. 
We attempt to avoid astrophysics modelling assumptions as much as possible. This is not an easy task, 
as much of the CR literature is focused on phenomenological modelling of propagation. We leave out most of the modelling questions\footnote{See~\cite{Strong:2007nh} for a useful review.}, hoping to provide a simple and robust understanding of CR antimatter that would be beneficial to particle physicists and astrophysicists alike. 

Our work is motivated by new observational information coming from an array of  experiments. Focusing on recent CR antimatter and closely related results we note, as a partial list of experimental contributions, the $e^\pm$ and $\pb$ measurements of PAMELA~\cite{Adriani:2008zr,Adriani:2010ib,Adriani:2012paa}, FERMI~\cite{Abdo:2009zk,FermiLAT:2011ab,Ackermann:2010ij}, ATIC~\cite{Chang:2008aa}, HESS~\cite{Aharonian:2008aa,Aharonian:2009ah}, and AMS02~\cite{PhysRevLett.110.141102,Aguilar:2016kjl,AMS02C2O:2016}. Future results are expected from CALET~\cite{Marrocchesi:2016jpn}, DAMPE~\cite{Wang:2016yov} and CTA~\cite{Vandenbroucke:2015gva}. Progress in the search for $\ad$ and $\ah$ is expected with AMS02~\cite{Giovacchini:2007dwa,kounineHebar}, GAPS~\cite{vonDoetinchem:2015zva,Aramaki:2015laa}, and BESS~\cite{Abe:2011nx,2012PhRvL.108m1301A}.

The layout of this review is as follows. 
In Sec.~\ref{sec:pbar} we consider secondary $\pb$ -- from the theoretical perspective, the simplest form of CR antimatter. 
In Sec.~\ref{ssec:grammage}-\ref{sec:pb2pb2c} we show how stable, relativistic, secondary nuclei data, under the general assumption that the CR {\it elemental composition}\footnote{Neither the over-all CR intensity, nor the target interstellar matter density, needs to be uniform in the propagation region in order for the procedure to apply.} is approximately uniform in the regions dominating spallation, allow one to calibrate away most of the propagation modelling uncertainties and extract a parameter-free prediction for $\pb$.  
The same calibration process is known to describe well the fluxes of secondary CR nuclei. For $\pb$, residual sensitivity remains to possible CR spectral variations in the spallation regions, as we discuss at some length. The insight behind this calibration process and a discussion of the conditions for its validity are presented in App.~\ref{s:gram}. We find the secondary $\pb$ prediction to be consistent with data within the uncertainties. 
In Sec.~\ref{ssec:pb2pmodels} we compare the model-independent analysis with certain models of propagation.

In Sec.~\ref{sec:ep} we turn to $\ep$, a hot potato: here public opinion basically has it that a primary source of $\ep$ must exist, be it dark matter or pulsars. 
We take a fresh look at the data in Sec.~\ref{s:pos}; the first thing we notice appears like a hint in the opposite direction: the observed $\ep/\pb$ flux ratio saturates the ratio of production rates in proton-proton collisions, a compelling hint for secondary $\ep$. 
We devote Sec.~\ref{sec:tetc} to elucidate the picture for $\ep$. If $\ep$ are secondary, then $\ep$ energy losses during propagation must be small, requiring that the CR propagation time is shorter than the time scale it takes $\ep$ to radiate a significant amount of their energy. Such a scenario cannot be accommodated in the conventional CR diffusion models~\cite{Strong:2007nh}, and this inconsistency with propagation models was the main cause for the claim of an ``$\ep$ anomaly"~\cite{Adriani:2008zr}. 
To be clear: we do not know of a fully satisfactory and tested alternative propagation model that reproduces the behaviour of $\ep$ with secondary production. However, putting modelling questions aside, we show in Sec.~\ref{sec:radnuc} that high energy radioactive nuclei data could test the secondary $\ep$ hypothesis in the near future. 
In Sec.~\ref{ssec:secepprop} we provisionally assume that $\ep$ are secondary to review some general lessons for CR propagation. These lessons which, again, are in tension with the currently common models of CR diffusion, may yet prove to be the long-term legacy of today's state of the art Galactic CR experiments. 
In Sec.~\ref{ssec:secep} we review some CR propagation models for secondary $\ep$. Finally, in 
Sec.~\ref{ssec:primep} we review ideas for primary $\ep$ or $\pb$ from pulsars and dark matter annihilation, showing that these models generically invoke fine-tuning to reproduce the observations.

In Sec.~\ref{sec:adah} we tackle the topic of CR $\ad$ and $\ah$. Surprisingly enough, we find a hot potato also here: we suggest, contrary to most earlier estimates, that a detection of secondary $\ah$ may be imminent at AMS02 (consistent with some pesky recent rumours).

In Sec.~\ref{sec:sum} we conclude. A short technical discussion of the interplay between $\ep$ radiative losses and propagation time is reserved to App.~\ref{app:demo}.

\section{Astrophysical $\pb$: the Galaxy as a fixed-target experiment}\label{sec:pbar}

CR antimatter particles are produced as secondaries in collisions of other CRs, notably protons, with interstellar matter (ISM), notably hydrogen in the Galaxy. Highly relativistic $\pb$ and heavier  antinuclei ($\ad,\,\ah$) propagate similarly to relativistic matter nuclei at the same magnetic rigidity\footnote{Here $p$ represents particle momentum, of course; elsewhere we often use the same symbol as shortcut for ``proton".}
\be\R=p/eZ,\no\ee 
with the difference in charge sign expected to make little or no impact on the propagation given that the measured CR flux is very nearly locally isotropic~\cite{DiSciascio:2014jwa}. 

Starting with the simplest case of $\pb$, it is natural to try and calibrate the effect of propagation directly from data, by using information on other secondary nuclei like boron (B), formed by fragmentation of heavier CRs (mostly carbon C and oxygen O). 
We now explain how to perform this calibration, calculate the predicted $\pb$ flux, and compare to measurements.

\subsection{The CR grammage}\label{ssec:grammage}

In this section we limit the discussion to stable, relativistic, secondary nuclei. For such secondaries, including e.g. B and the sub-Fe group (T-Sc-V-Cr), the ratio of densities of two species $a,b$ satisfies an approximate empirical relation~\cite{Engelmann:1990zz,2003ApJ...599..582W},
\be\label{eq:sec}\frac{n_a(\R)}{n_b(\R)}&\approx&\frac{Q_a(\R)}{Q_b(\R)}.\ee
Here $Q_a$ denotes the net production of species $a$ per unit ISM column density, 
\be\label{eq:Q} Q_a(\R)&=&\sum_Pn_P(\R)\frac{\sigma_{P\to a}(\R)}{m}-n_a(\R)\frac{\sigma_a(\R)}{m},\ee
where $(\sigma_a/m)$ and $(\sigma_{P\to a}/m)$ are the total inelastic and the partial $P\to a$ cross section per target ISM particle mass $m$, respectively. 

We stress that Eq.~(\ref{eq:sec}) is an empirical relation, known to apply to $\sim$10\% accuracy in analyses of HEAO3 data~\cite{Engelmann:1990zz,2003ApJ...599..582W} and -- as we shall see shortly, focusing on $\pb$ -- consistent with subsequent PAMELA~\cite{Adriani:2012paa} and AMS02~\cite{Aguilar:2016kjl} measurements. 
In applying Eq.~(\ref{eq:sec}) to $\pb$, a subtlety arises due to the fact that the cross sections appearing in Eq.~(\ref{eq:Q}) can (and for $\pb$, do) depend on energy. In Eq.~(\ref{eq:Q}) we define these cross sections such that the source term $Q_a(\R)$ is proportional to the progenitor species density $n_P(\R)$ expressed at the same rigidity; we will clarify this statement further down the road in Eq.~(\ref{eq:pbcs}). For relativistic nuclei (above a few GeV/nuc) produced in fragmentation reactions, e.g. $^{12}$C fragmenting to $^{11}$B, the energy dependence of the fragmentation cross section is much less important. 

From the theoretical point of view, Eq.~(\ref{eq:sec}) is natural~\cite{Katz:2009yd,1996ApJ...465..972P,1998ApJ...505..266S,Ginzburg:1990sk}. It is guaranteed to apply if the relative {\it elemental composition} of the CRs in the regions that dominate the spallation is similar to that measured locally at the solar system\footnote{Note: neither the over-all CR intensity, nor the target ISM density, needs to be uniform in the propagation region in order for Eq.~(\ref{eq:sec}) to apply. Indeed, the ISM exhibits orders of magnitude variations in density across the Galactic gas disc and rarified halo~\cite{Ferriere:2001rg}.}: in this case, the source distribution of different secondaries is similar. 
Because the confinement of CRs in the Galaxy is magnetic, different CR particles that share a common distribution of sources should exhibit similar propagation if sampled at the same rigidity\footnote{This is, of course, provided that the CR species being compared do not exhibit species-dependent complications like decay in flight (for radionuclei like $^{10}$Be) or radiative energy losses (for $\ep$). In addition, rigidity only really becomes the magic quantity for propagation at relativistic energies (see e.g.~\cite{2003ApJ...599..582W}).}. Thus, the ratio of propagated CR densities reflects the ratio of their net production rates. 

Note that the net source defined in Eq.~(\ref{eq:Q}) accounts for the fact that different nuclei exhibit different degree of fragmentation losses during propagation. In this way, species like sub-Fe (with fragmentation loss cross section of order 500~mb), B ($\sigma_B\sim240$~mb), and $\bar p$ ($\sigma_{\pb}\sim40$~mb) can be put on equal footing.

Further discussion of the physical significance of Eq.~(\ref{eq:sec}) is given in Ref.~\cite{Katz:2009yd} and App.~\ref{s:gram}.\\

We can use Eq.~(\ref{eq:sec}) together with the locally measured flux of B, C,  O, p, He,... to predict the $\bar p$ flux~\cite{1992ApJ...394..174G,Katz:2009yd}:
\be\label{eq:pbfromB} n_{\bar p}(\R)&\approx&\frac{n_{\rm B}(\R)}{Q_{\rm B}(\R)}Q_{\bar p}(\R).\ee
The RHS of Eq.~(\ref{eq:pbfromB}) is derived from laboratory cross section data and from direct local measurements of CR densities, without reference to the details of propagation.

The quantity 
\be\label{eq:Xbc} \X(\R)&=&\frac{n_{\rm B}(\R)}{Q_{\rm B}(\R)},\ee
known as the CR grammage~\cite{Ginzburg:1990sk}, is a spallation-weighted average of the column density of ISM traversed by CRs during their propagation, the average being taken over the ensemble of propagation paths from the CR production regions to Earth. Combining AMS02 B/C~\cite{Aguilar:2016vqr} and C/O~\cite{AMS02C2O:2016} 
with heavier CR data from HEAO3~\cite{Engelmann:1990zz} and with laboratory fragmentation cross section data~\cite{Blum:2017qnn,Tomassetti:2015nha}, one can derive $\X$ directly from measurements:
\be\label{eq:X}\X&=&\frac{\rm (B/C)}{\sum_{\rm P=C,N,O,...}{\rm(P/C)}\frac{\sigma_{\rm P\to B}}{m}-{\rm(B/C)}\frac{\sigma_{\rm B}}{m}}.
\ee

\begin{figure}[t]
\begin{center}
\includegraphics[scale=0.4]{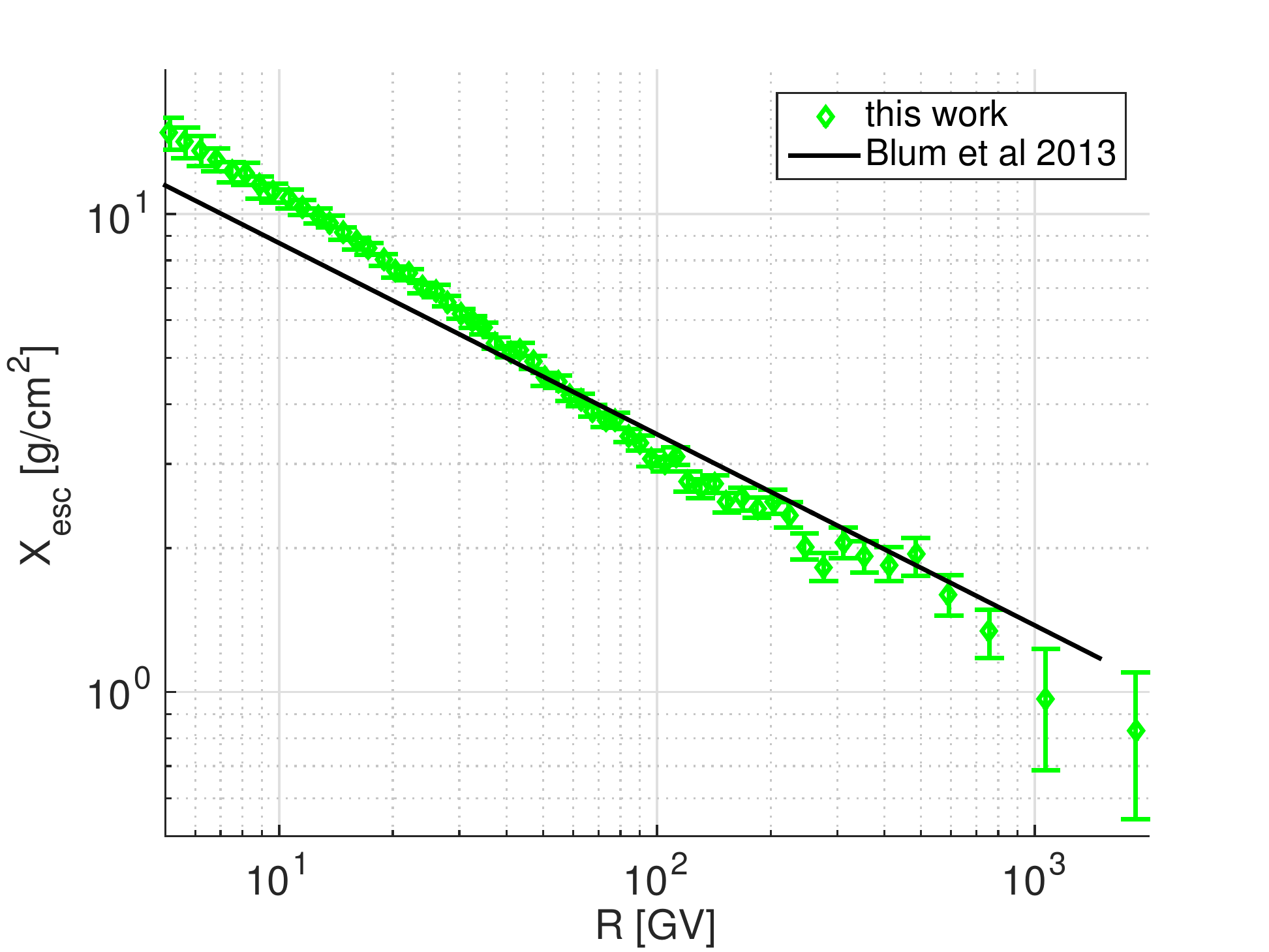}\quad
\includegraphics[scale=0.4]{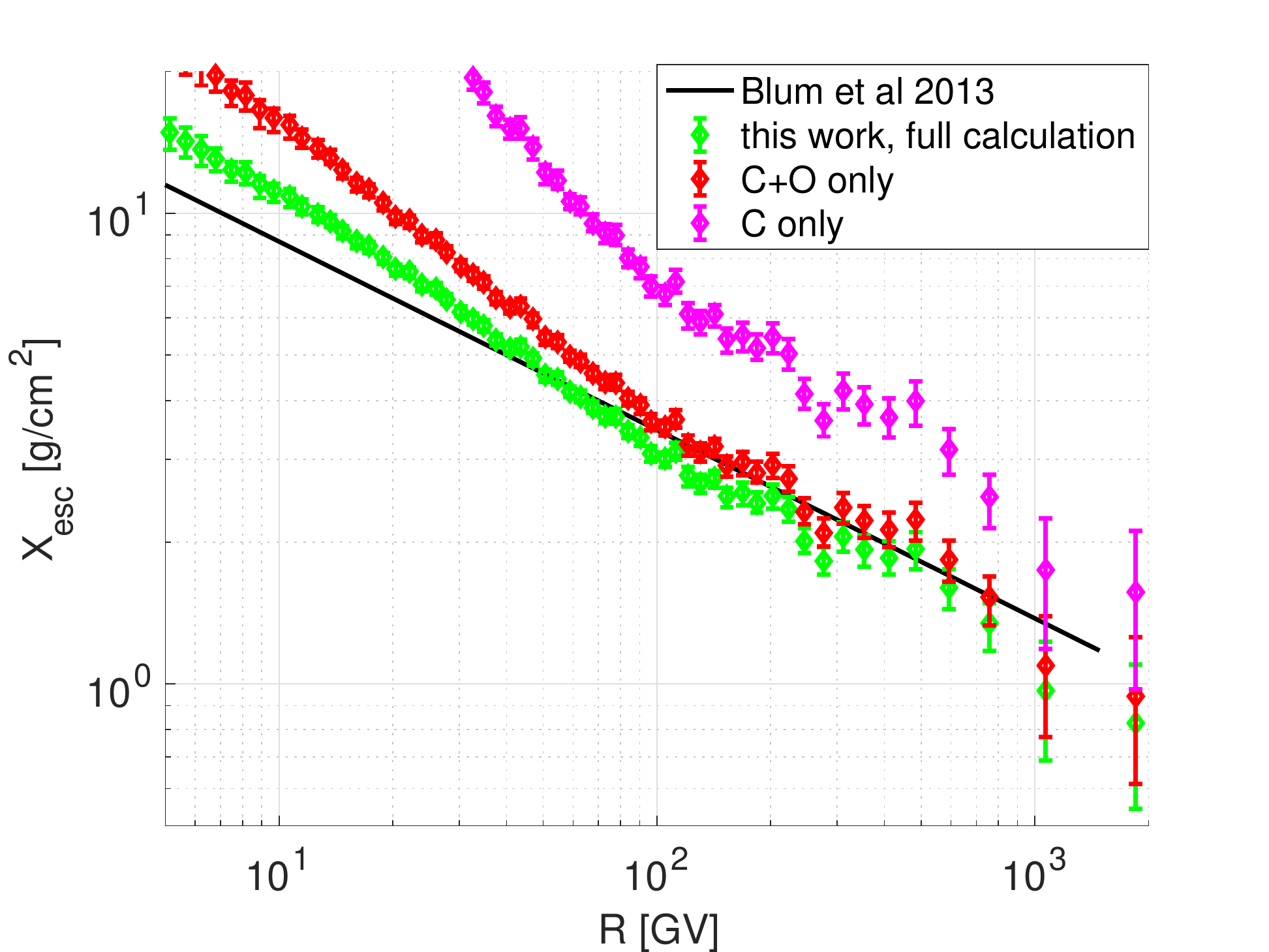}
\caption{{\bf Left:} CR grammage $\Xe$ derived directly from B/C, C/O, and heavier nuclei data and compared with the earlier approximation of~\cite{Blum:2013zsa}. {\bf Right:} separating various contributions to the full $\Xe$ result. Error bars represent only the B/C error bars reported in~\cite{Aguilar:2016vqr}, and do not include systematic uncertainties on fragmentation cross sections and on the flux ratios.}
\label{fig:X}
\end{center}
\end{figure}

The result for $\X$ is shown by the green markers in the left panel of Fig.~\ref{fig:X}. Error bars reflect the B/C error bars reported in~\cite{Aguilar:2016vqr}, and do not include systematic uncertainties on fragmentation cross sections and on the flux ratios C/O, N/O, etc. We estimate that the systematic fragmentation cross section uncertainties are at the level of 20\%; note that many of the cross sections used in the analysis at high energy are extrapolated from much lower energy data, typically confined to a few GeV/nuc. The result in Fig.~\ref{fig:X} agrees with the power-law approximation derived in Ref.~\cite{Blum:2013zsa} to 20\% accuracy.

To exhibit the different contributions entering the determination of $\Xe$, in the right panel of Fig.~\ref{fig:X} we show the result for $\Xe$ that obtains if we omit, in the B production source, the contributions due to all CR species other than C (purple markers), all species other than C+O (red markers).

\subsection{$\pb/p$ from B/C}\label{sec:pb2pb2c}

Now that we have $\X$, we can use the $\bar p$ production and loss cross sections parametrised in, e.g.,~\cite{1983JPhG....9..227T,0305-4616-9-10-015,Winkler:2017xor} together with measurements of the proton and helium~\cite{Aguilar:2015ooa,Aguilar:2015ctt,0004-637X-839-1-5} flux to calculate $Q_{\bar p}$ and apply it in Eq.~(\ref{eq:pbfromB}). At low rigidity, the effect of solar modulation is estimated as in~\cite{2003ApJ...599..582W} with $\Phi=450$~MV. 

\begin{figure}[t]
\begin{center}
\includegraphics[scale=0.6]{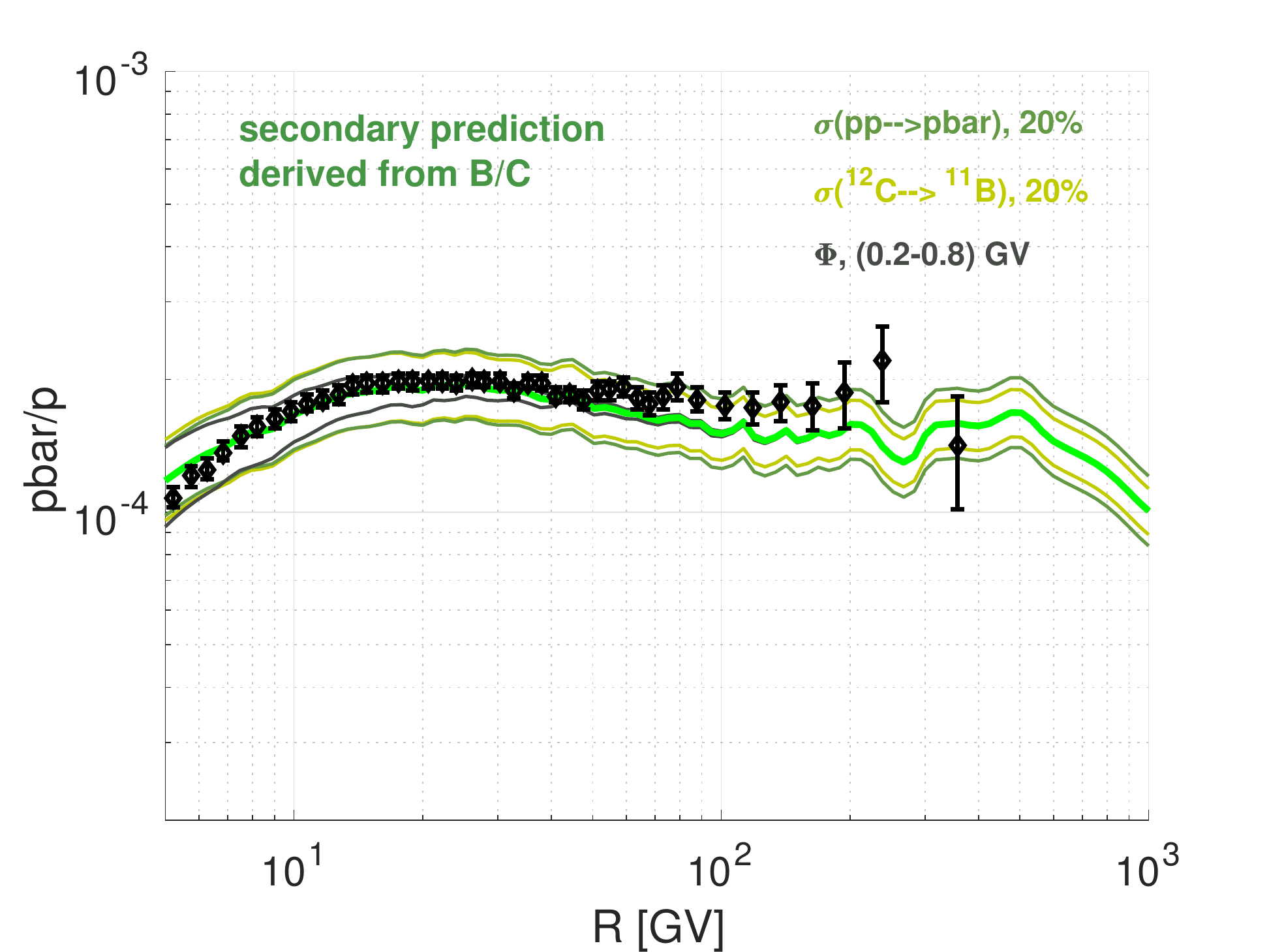}
\caption{
Observed $\bar p/p$ ratio~\cite{Aguilar:2016kjl} vs. the secondary prediction, calculated using the locally measured B/C ratio and p and He flux. Wiggles in the theory curve come from our direct data-driven use of the CR grammage, and reflect fluctuations in the AMS02 B/C data~\cite{Aguilar:2016vqr}. Thick line is the secondary prediction with input cross sections detailed in~\cite{Blum:2017qnn}, while thin lines show the response of the prediction for variation in (i) $pp\to\bar p$ cross section within $\pm20\%$, (ii) fragmentation cross section p$^{12}$C$\to$$^{11}$B within $\pm20\%$, (iii) variation in the solar modulation parameter in the range $\Phi=(0.2-0.8)$~GV. Taken from Ref.~\cite{Blum:2017qnn}. }
\label{fig:pb}
\end{center}
\end{figure}

The result is compared to AMS02 data~\cite{Aguilar:2016kjl,Nozzoli:2016ske} in Fig.~\ref{fig:pb}. The $\pb$ flux is consistent, within statistical and systematic uncertainties, with the prediction of Eq.~(\ref{eq:pbfromB}). No astrophysical propagation modelling is needed: Eq.~(\ref{eq:pbfromB}) has successfully calibrated out propagation from B/C data. 

We can conclude that CR $\pb$ are most likely secondary. 

As mentioned earlier, in computing $Q_{\pb}$ we need to account for the energy-dependent $\pb$ production cross section. Let us consider the main positive contribution to $Q_{\pb}$, due to pp collisions. Because of a kinematical barrier, the daughter $\pb$ is emitted with a rigidity smaller by a factor of $\sim10$ compared to the rigidity of the parent $p$ in the ISM frame. Given a spectrum of parent protons, one can still compute the overlap integral between the differential cross section and the parent p spectrum and express the contribution to $Q_{\bar p}$ in terms of an effective cross section $\sigma_{p\to\bar p}(\R)$:
\be\label{eq:pbcs}\sigma_{p\to\bar p}(\R)&=&\frac{2\int_{\R}^\infty d\R_p\,J_p(\R_p)\left(\frac{d\sigma_{pp\to\bar pX}(\R_p,\R)}{d\R_p}\right)}{J_p(\R)}.\ee
The factor of 2 above\footnote{See~\cite{Winkler:2017xor} for a recent examination of isospin asymmetry in $pp\to\pb,\bar n$.} accounts for the production and subsequent decay of $\bar n$, with $\sigma_{pp\to\bar nX}\approx\sigma_{pp\to\bar pX}$. A similar procedure is used to include the contributions due to proton CR hitting He in the ISM; He CR hitting ISM hydrogen; and so forth. 

Calculating Eq.~(\ref{eq:pbcs}) for a power-law proton flux $J_p\propto\R^{-\gamma_p+\Delta\gamma_p}$, one finds the scaling $\sigma_{p\to\bar p}^{(\Delta\gamma_p)}\approx10^{\Delta\gamma_p}\,\sigma_{p\to\bar p}^{(\Delta\gamma_p=0)}$~\cite{Katz:2009yd}. 
The effective cross section $\sigma_{p\to\bar p}(\R)$ therefore depends on the spectral shape of CR protons. This is a new feature compared to the heavy nuclei fragmentation cross sections: there, due to the straight-ahead kinematics, a cross section like $\sigma_{\rm C\to B}$ is independent of the carbon spectral index to a good approximation. 

Calculating $\sigma_{p\to\bar p}(\R)$ with the locally measured p flux, one might expect deviations from Eq.~(\ref{eq:sec}) if the proton spectral shape in the spallation regions exhibits variations compared to its locally measured value. Models that realise this possibility include~\cite{Cowsik:2009ga,Burch:2010ye,Cowsik:2013woa,Blasi:2009hv,Blasi:2009bd,Mertsch:2009ph,Ahlers:2009ae,Kachelriess:2011qv,Cholis:2017qlb,Kachelriess:2015oua,Thomas:2016fcp}, reviewed in Sec.~\ref{ssec:secep}.

\begin{figure}[t]
\begin{center}
\includegraphics[scale=0.55]{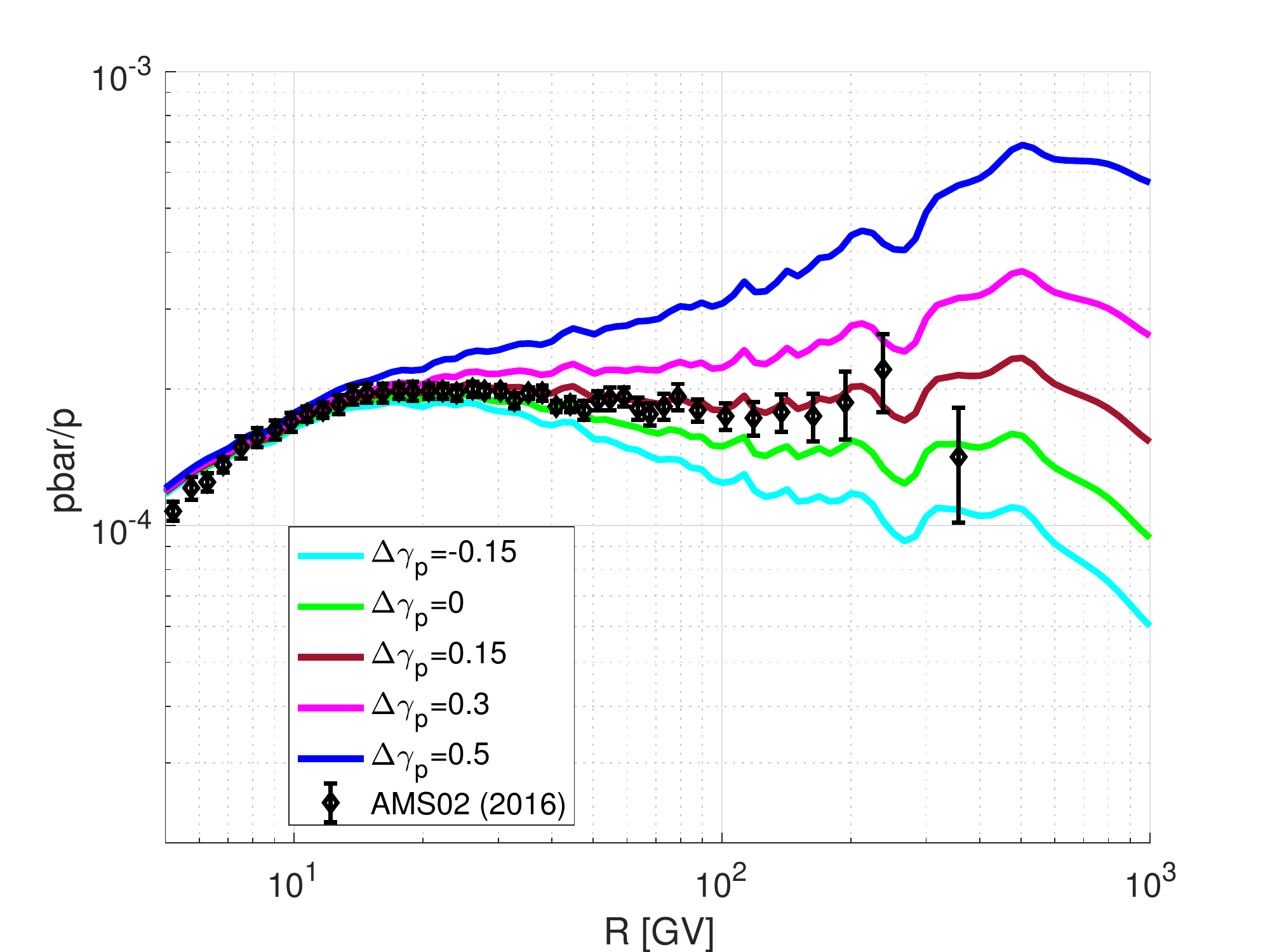}
\caption{Same as Fig.~\ref{fig:pb}, but showing in cyan, green, red, magenta, and blue the resulting local value of $\pb/p$ that obtains if the high energy proton spectrum in the spallation region is changed w.r.t. the locally measured spectrum by a factor $\left(\R/200~{\rm GV}\right)^{\Delta\gamma_p}$, starting at $\R>200$~GV, with $\Delta\gamma_p=-0.15,\,0,\,0.15,\,0.3,\,0.5$, respectively.}
\label{fig:pb2}
\end{center}
\end{figure}

In Fig.~\ref{fig:pb2} we quantify the sensitivity of the locally measured $\pb/p$ ratio to a difference between the primary CR spectrum measured locally, to the spectrum in the regions of the Galaxy that dominate the secondary production. Similarly, the plot also exhibits the sensitivity in the predicted $\pb/p$ ratio to measurement errors in determining the local proton flux\footnote{The systematic difference between different experimental determinations of the local CR proton and He flux is not negligible, at a level of 10-20\% with larger extrapolation uncertainty in the relevant few TV range; see e.g.~\cite{Panov:2011ak,Adriani:2011cu,Adriani:2013xva,Aguilar:2015ooa,Aguilar:2015ctt,0004-637X-839-1-5}.}. We show the $\pb/p$ ratio that obtains if, in the calculation of $\sigma_{p\to\pb}(\R)$, we replace the locally measured proton flux by a modified spectrum: $J_{p}(\R)\to J_{p}(\R)F_p(\R)$, with $F_P(\R)=1$ for $\R\leq200$~GV and $F_p(\R)=\left(\R/{\rm 200~GV}\right)^{\Delta\gamma_p}$ for $\R>200$~GV. The result is shown for the choices $\Delta\gamma_p=-0.15,\,0,\,0.15,\,0.3,\,0.5$. 
Note that our parametrization of the locally measured proton flux uses a direct (non power-law) interpolation of AMS02~\cite{Aguilar:2015ooa} and CRAM-III~\cite{0004-637X-839-1-5} data.

Allowing room for cross section and CR spectra uncertainties, from inspection of Fig.~\ref{fig:pb2} we infer a rough limit: 
\be\label{eq:Dgp}
-0.15<\Delta\gamma_p<0.4.\ee 
Eq.~(\ref{eq:Dgp}) applies to the situation where all primary CR spectra are modified in the secondary production regions, compared to their local value; namely, the variation in proton and in C and O spectra is correlated, $\Delta\gamma_{\rm C,O}\approx\Delta\gamma_p$. This could occur with nontrivial propagation from the CR fragmentation regions to our local spot in the Milky Way. 
One may also constrain the possibility of CR accelerators that inject a different composition of primary CR spectra in different (but connected by propagation) regions of the Galaxy, in which case $\Delta\gamma_{\rm C,O}\neq\Delta\gamma_p$. This situation corresponds to significantly non-uniform CR composition in the propagation region, invalidating the grammage relation. While we do not pursue this analysis here, it is also constrained by the success of Eq.~(\ref{eq:pbfromB}) in reproducing the $\pb/p$ data. 
\\
 
To conclude so far: 
\begin{itemize}
\item When it comes to relativistic, stable, secondary nuclei and antinuclei, the Galaxy is essentially a fixed-target experiment, with CRs themselves playing the role of the beam and with ISM being the target. This simple point and the resulting predictions are often obscured in calculations done within specific models of CR propagation. 
\item Fig.~\ref{fig:pb} demonstrates that the $\bar p$ flux measured by AMS02~\cite{Aguilar:2016kjl} is consistent with secondary production, within current fragmentation and $pp\to\pb$ cross section uncertainties and the uncertainties in B/C and primary CR spectra. 
\end{itemize} 

It is worthwhile to compare our analysis to results obtained in the context of specific models of propagation. Of these, the most commonly used are the two-zone disc+halo diffusion models~\cite{Strong:2007nh}. In the next section we discuss these models in view of the $\pb$ data.

\subsection{$\pb$ in diffusion models}\label{ssec:pb2pmodels}

Typical two-zone diffusion models in the literature satisfy the simple condition leading to Eq.~(\ref{eq:sec}), so they too satisfy Eq.~(\ref{eq:sec}), at least approximately~\cite{Ginzburg:1990sk}. This is because the relative composition of the CRs in these models is approximately uniform across the thin ISM disc, where most of the spallation happens.  
It is interesting to compare results obtained within these models to results derived directly from Eq.~(\ref{eq:sec}).

\begin{figure}[t]
\begin{center}
\includegraphics[scale=0.6]{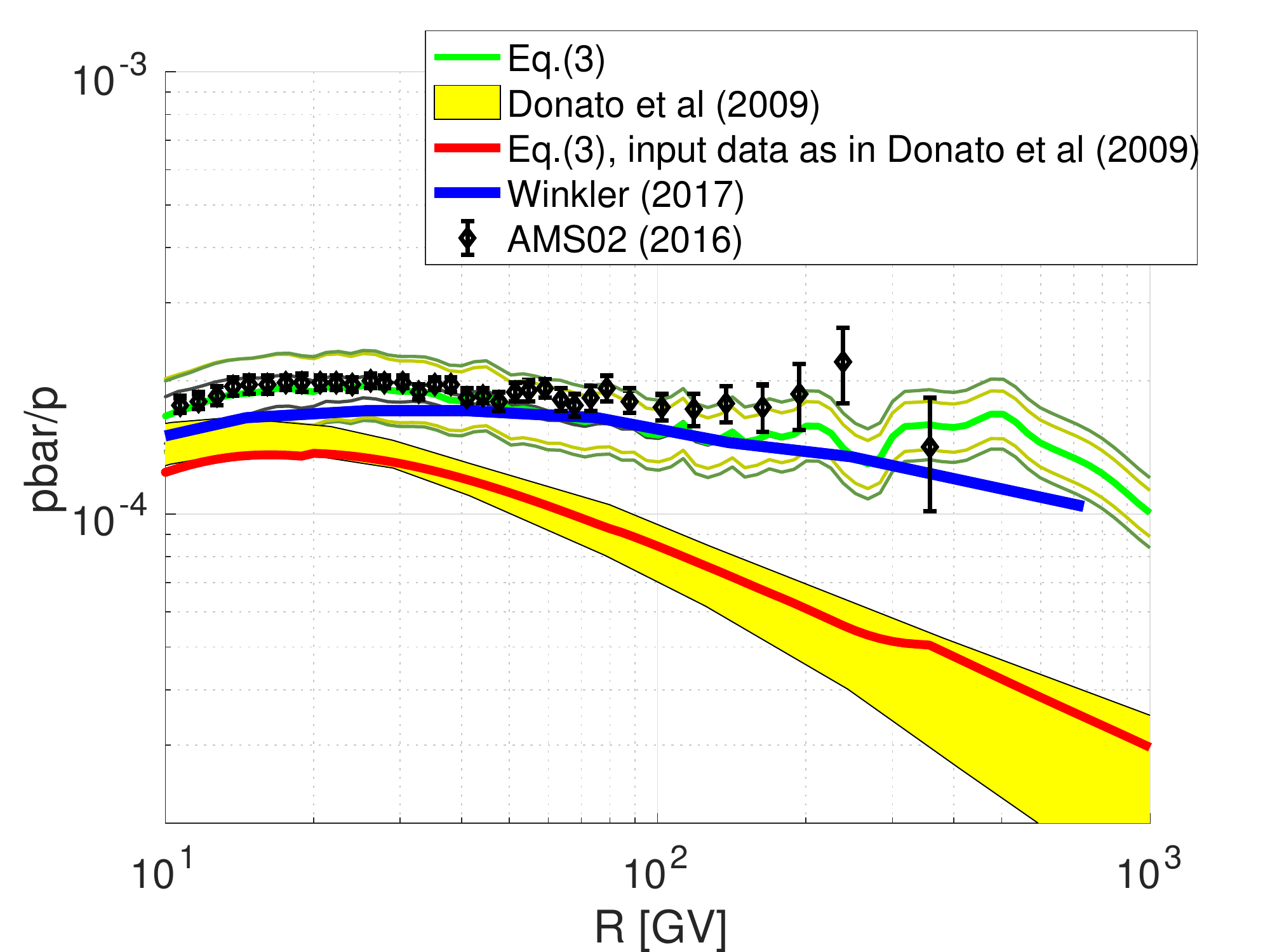}
\caption{
Comparison of the result of Eq.~(\ref{eq:pbfromB}) (green) with results from the diffusion models of~\cite{Winkler:2017xor} (blue) and~\cite{Donato:2008jk} (yellow band). AMS02 data in black.  To understand the discrepancy with~\cite{Donato:2008jk}, we use the model parameters of Ref.~\cite{Donato:2008jk}, based on early HEAO3 B/C data, to compute the effective $\Xe$. We then use this $\Xe$ to calculate the $\pb$ flux. We also adopt the primary p and He spectra and $pp\to\bar p$ cross section parametrisation of~\cite{Donato:2008jk}. The result we obtain in this way is shown in red. 
}
\label{fig:Donato09}
\end{center}
\end{figure}

A recent example of the diffusion model was given in~\cite{Kappl:2015bqa,Winkler:2017xor}. Ref.~\cite{Kappl:2015bqa,Winkler:2017xor} calibrated the diffusion model to AMS02 B/C data, and used the resulting model parameters together with state of the art proton and He data to calculate the $\pb$ flux. The result, taken from~\cite{Winkler:2017xor}, is shown by the blue line in Fig.~\ref{fig:Donato09}. As can be seen, this result is consistent with the result obtained directly from Eq.~(\ref{eq:sec}), using the same B/C, p and He data and a similar $pp\to\pb$ cross section code.

As another example, Ref.~\cite{Donato:2008jk} used a diffusion model with the same set of assumptions, geometry, and free parameters to that of~\cite{Kappl:2015bqa,Winkler:2017xor}. However, the $\pb/p$ ratio predicted in~\cite{Donato:2008jk}, shown by the yellow band in Fig.~\ref{fig:Donato09}, falls significantly below the AMS02 $\pb/p$ data. What went wrong?

The main thing that went wrong, is that the model of Ref.~\cite{Donato:2008jk} was calibrated to fit early B/C data from HEAO3~\cite{Engelmann:1990zz}, and then extrapolated from that fit to high energy beyond the region where HEAO3 data was tested. Unfortunately, above $\R\sim100$~GV the extrapolation of the HEAO3 B/C data falls bellow the more recent AMS02 measurement. 
In addition, \cite{Donato:2008jk} assumed a primary proton flux with high energy spectral index $\gamma_p=2.84$, softer by about $\Delta\gamma_p\approx-0.12$ than the proton flux seen by AMS02. The implications of this soft high energy proton spectrum can be estimated from Fig.~\ref{fig:pb2}. At the same time, at lower energy $\R\lesssim100$~GV the assumed proton flux was higher than that reported by AMS02, decreasing further the predicted $\pb/p$ ratio.

As a result, Ref.~\cite{Donato:2008jk} predicted a low $\pb/p$ ratio. To illustrate this fact, we show in Fig.~\ref{fig:Donato09} by a red line the result we find if we calculate the $\pb/p$ ratio using Eq.~(\ref{eq:sec}), but using the value of $\Xe$ derived for the diffusion model of~\cite{Donato:2008jk} with $\delta=0.7$ and using the same proton flux assumed there\footnote{We use Eq.~(\ref{eq:Xediff}) to calculate $\Xe$ in the diffusion model, with a thin disc thickness of 100~pc and disc ISM proton density of 1~cm$^{-3}$. More details on the diffusion model can be found in Sec.~\ref{sec:tetc}.}. The discrepancy with Ref.~\cite{Donato:2008jk} is reproduced. 

\section{What is the issue with $e^+$?}\label{sec:ep}

Measurements of the positron fraction $\ep/e^\pm=\ep/(\ep+e^-)$ by the PAMELA~\cite{Adriani:2008zr,Adriani:2010ib} and AMS02~\cite{PhysRevLett.110.141102} experiments have shown that $\ep/e^\pm$ is rising with energy from a few GeV to at least 300~GeV. This trend of rising $\ep/e^\pm$ was claimed by many to indicate a primary source dominating the $\ep$ flux at these energies. Understanding the true story behind CR $\ep$ is crucial: models for a primary source range from dark matter annihilation\footnote{See~\cite{Bergstrom:2008gr,Cholis:2008hb,Barger:2008su,Cirelli:2008jk} as representative examples.} to a contribution of $e^\pm$ from pulsars\footnote{See~\cite{PhysRevD.52.3265,1989ApJ...342..807B,Profumo:2008ms,Hooper:2008kg} as representative examples.}, both exciting ideas.

Our goal in this section is to consider what can be learned from the $\ep$ measurements. 
We start with inspecting the data, in Sec.~\ref{s:pos}, pointing out that the observed $\ep/\pb$ flux ratio saturates the ratio of production rates $Q_{\ep}/Q_{\pb}$ in proton-proton collisions. This is a compelling hint for secondary $\ep$. 

What then is the basis to the claim that a primary $\ep$ source is required? 
In Sec.~\ref{sec:tetc} we show that if $\ep$ are secondary, then $\ep$ energy losses during propagation must be small, requiring that the CR propagation time is shorter than the time scale it takes $\ep$ to radiate a significant amount of their initial energy (of order $\sim1$~Myr at $\R\sim300$~GV). Such a scenario cannot be accommodated in the conventional phenomenological diffusion models~\cite{Strong:2007nh}, and the inconsistency with propagation models was the main cause for the claim of an ``$\ep$ anomaly"~\cite{Adriani:2008zr}. However, we know of no contradiction of this scenario with either observational data or first principle theory. 

The nearest complimentary data that could test the secondary $\ep$ hypothesis involves radioactive secondary nuclei, and we review it in Sec.~\ref{sec:radnuc}. In Sec.~\ref{ssec:secepprop} we entertain the possibility that $\ep$ are indeed secondary, and deduce some basic model-independent lessons for CR astrophysics. In Sec.~\ref{ssec:secep} we review some CR propagation models for secondary $\ep$. Finally, in 
Sec.~\ref{ssec:primep} we review ideas for primary $\ep$ or $\pb$ from pulsars and dark matter annihilation, showing that these models generically invoke fine-tuning to reproduce the observations.

\subsection{A hint for secondary $\ep$}\label{s:pos}

In dealing with $\pb$ we used Eq.~(\ref{eq:sec}), cast in the form of Eq.~(\ref{eq:pbfromB}) where $\Xe=n_B/Q_B$ is derived from nuclei data. However, Eq.~(\ref{eq:sec}) cannot be directly applied to predict the flux of $\ep$, because $\ep$ are subject to radiative energy losses and Eq.~(\ref{eq:sec}) does not capture the effect of energy loss during propagation
(see discussion in App.~\ref{s:gram}).

Nevertheless, we can still gain insight from Eq.~(\ref{eq:sec}). As noted in~\cite{Katz:2009yd},  Eq.~(\ref{eq:sec}), applied to $\ep$ with radiative losses ignored, provides an upper limit to the secondary $\ep$ flux because radiative energy losses can only decrease\footnote{This statement is true for a steeply falling $\ep$ spectrum, assuming synchrotron and inverse-Compton (IC) losses dominate in the Thomson regime. The requirement to avoid pile-up in this case is $\gamma_{\ep}>2$, to be compared with the observed $\gamma_{\ep}\sim2.75$.} the secondary $\ep$ flux compared to the loss-less secondary benchmark. This provides an {\it upper bound} on the flux of secondary $\ep$.

The most robust way to formulate the secondary upper bound on the $\ep$ flux is in terms of branching fractions in pp collisions, comparing $\ep$ to $\pb$. The upper bound reads:
\be\label{eq:pb2pos}
\frac{n_{\ep}}{n_{\pb}}&\leq&\frac{Q_{e^+}(\R)}{Q_{\bar p}(\R)}.\ee

\begin{figure}[t]
\begin{center}
\includegraphics[scale=0.7]{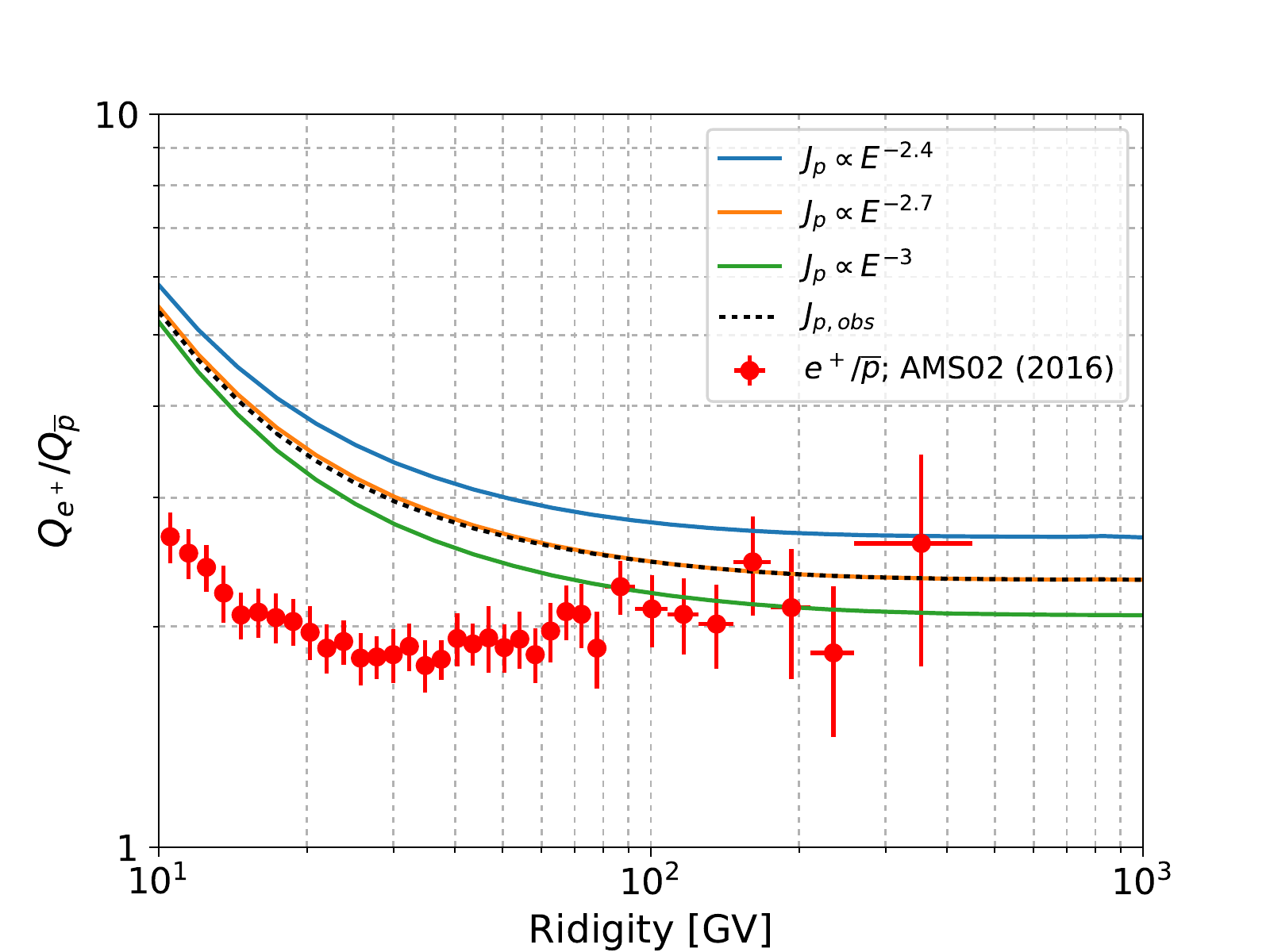}
\caption{$\ep/\pb$ flux ratio: AMS02 data compared to the secondary upper bound of Eq.~(\ref{eq:pb2pos}). The upper bound ($\ep/\pb$ source ratio) is shown with different assumptions for the proton spectrum in the secondary production regions. Systematic cross section uncertainties in $pp\to\pb,\ep$, not shown in the plot, are in the ballpark of 10\%. Dashed black line shows the result evaluated for the locally measured $J_p$, while blue and green lines show the result for harder and softer proton flux, respectively, as specified in the legend. Taken from~\cite{Blum:2017iol}.}
\label{fig:pb2pos}
\end{center}
\end{figure}

AMS02 have recently reported the inverse ratio $\bar p/e^+$~\cite{Aguilar:2016kjl}. The experimental results are compared to the bound of Eq.~(\ref{eq:pb2pos}) in Fig.~\ref{fig:pb2pos}.
\begin{figure}[t]
\begin{center}
\includegraphics[scale=0.4]{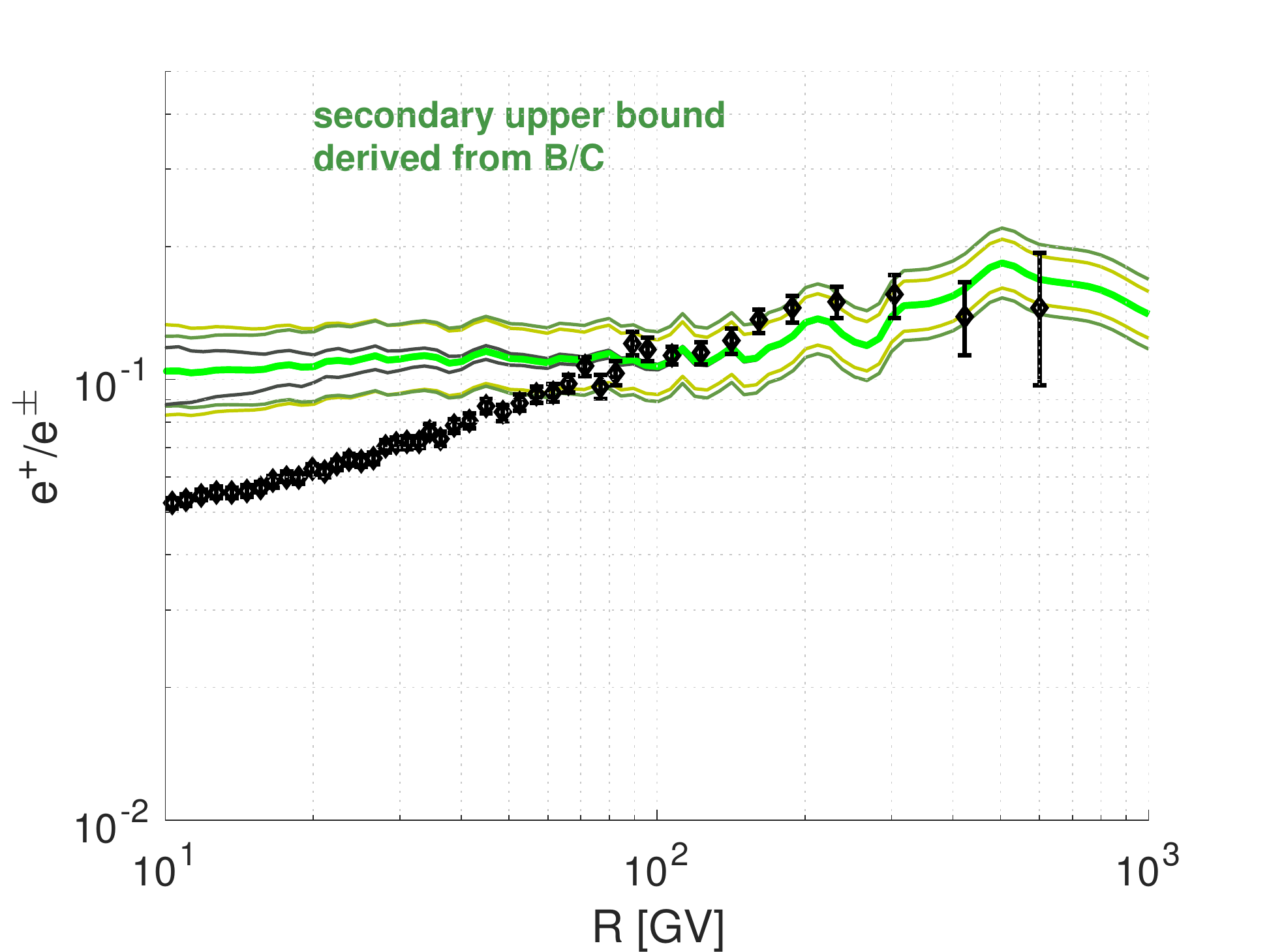}\quad
\includegraphics[scale=0.4]{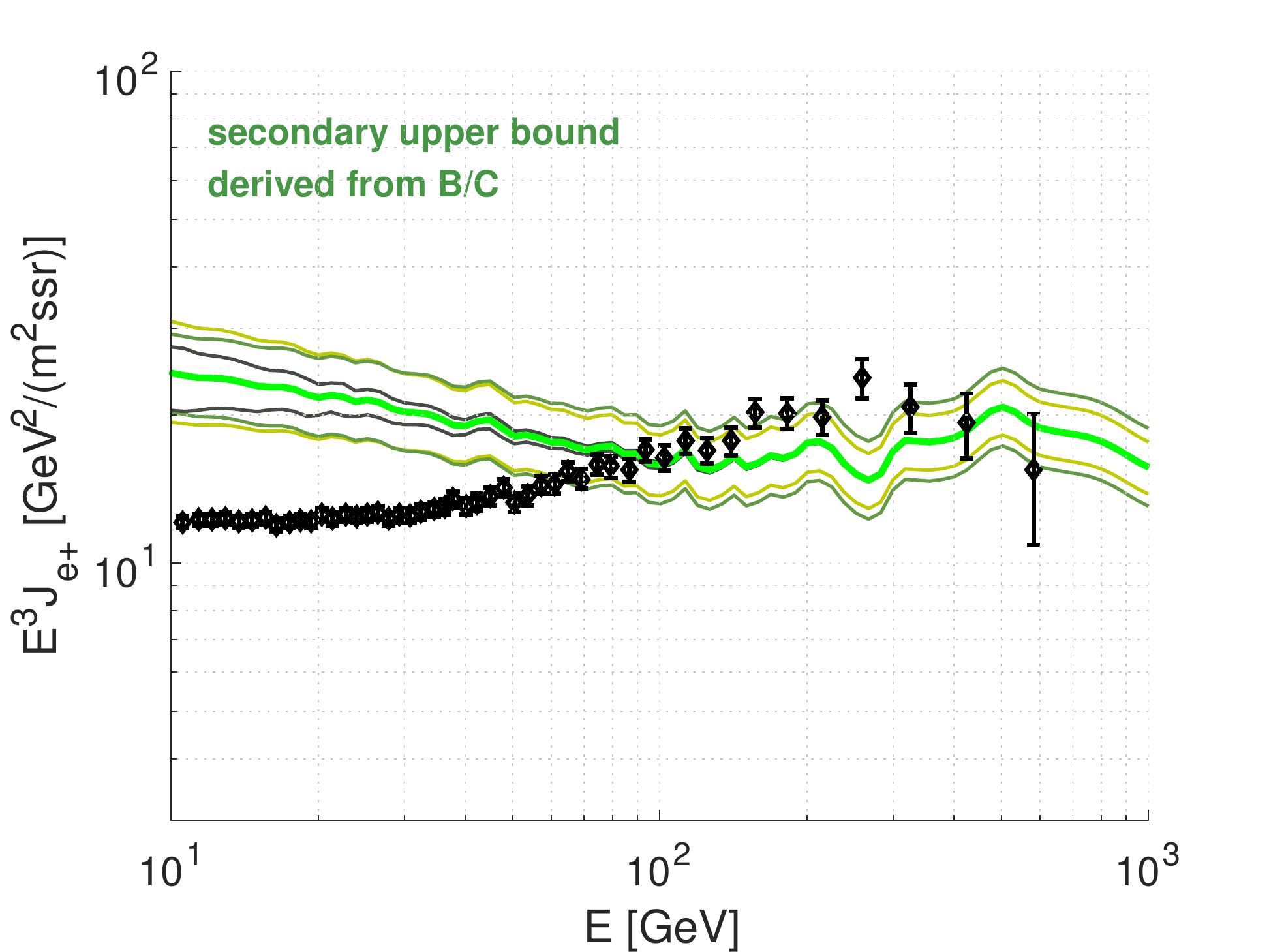}
\caption{{\bf Left:} $\ep/e^{\pm}$ flux ratio. AMS02 data compared to the secondary upper bound, evaluated directly from B/C data by using the equivalent form of Eq.~(\ref{eq:pbfromB}) applied to $\ep$. Systematic uncertainties are represented as in Fig.~\ref{fig:pb}. {\bf Right:} same as on the left but showing the $\ep$ flux.}
\label{fig:posfrac}
\end{center}
\end{figure}
\begin{figure}[t]
\begin{center}
\includegraphics[scale=0.55]{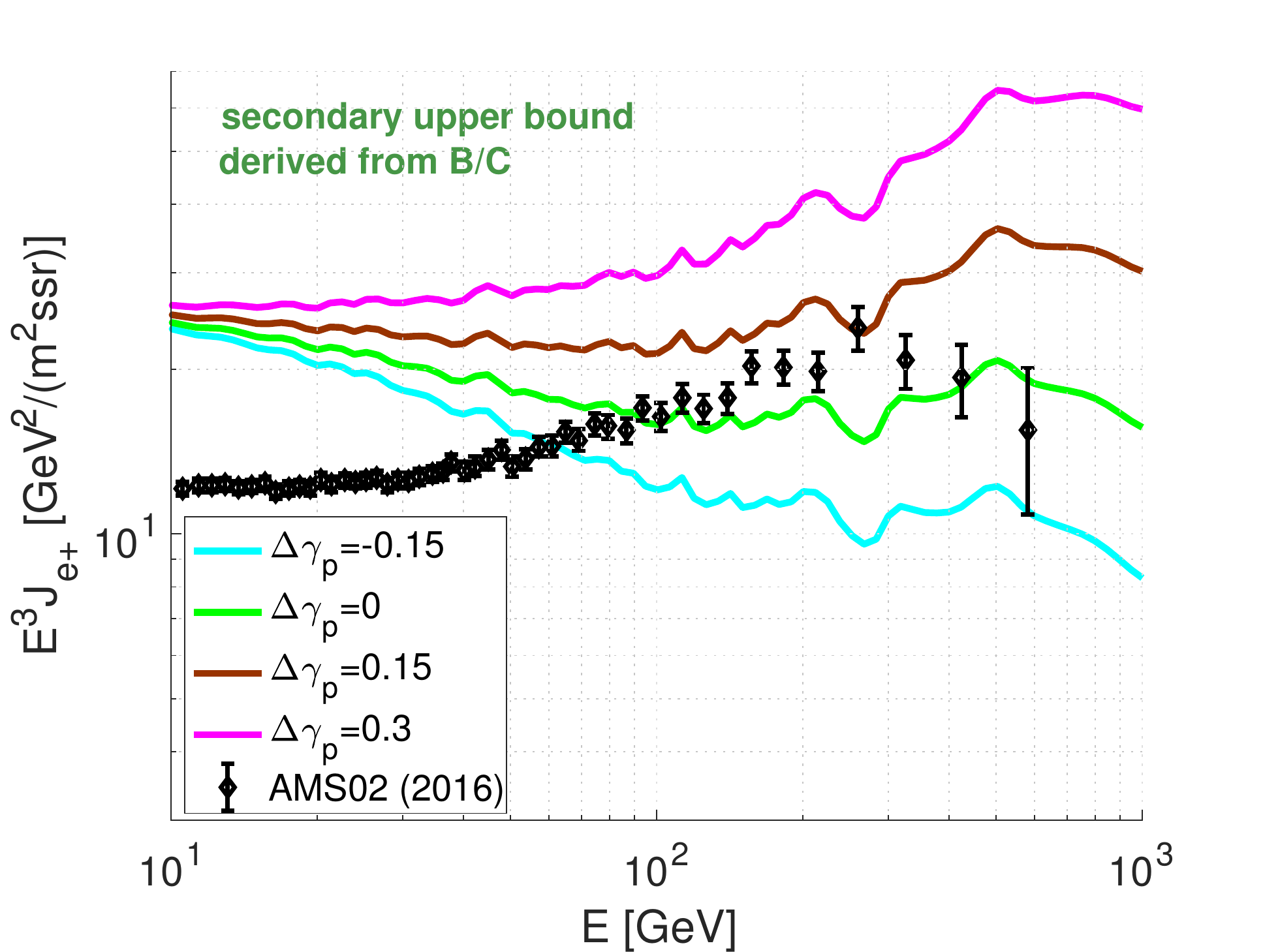}
\caption{Same as the right panel of Fig.~\ref{fig:posfrac}, but showing in cyan, green, red, and magenta the resulting secondary upper limit on $J_{\ep}$ that obtains if the high energy proton spectrum in the spallation region is changed w.r.t. the locally measured spectrum by a factor $\left(\R/200~{\rm GV}\right)^{\Delta\gamma_p}$, starting at $\R>200$~GV, with $\Delta\gamma_p=-0.15,\,0,\,0.15,\,0.3$, respectively.}
\label{fig:pos2}
\end{center}
\end{figure}

Pause to appreciate this situation: The measured $\ep/\pb$ ratio does not exceed and is always comparable to the secondary upper bound. Moreover, the $\ep/\pb$ ratio saturates the bound over an extended range in rigidity. Taking into account that, as we saw in the previous section, $\pb$ are likely secondary (certainly dominated by secondary production), it is natural to deduce that {\bf AMS02 is observing secondary $\ep$} as well, and propagation energy losses are small.

A compatible but less robust way to represent the secondary $\ep$ upper bound is directly from the B/C grammage, as we did for $\pb$ in Fig.~\ref{fig:pb}. Namely, we write
\be\label{eq:posfromB} n_{\ep}(\R)&\lesssim&\frac{n_{\rm B}(\R)}{Q_{\rm B}(\R)}Q_{\ep}(\R).\ee
The result is shown in Fig.~\ref{fig:posfrac}. On the left panel the measured total $e^\pm$ flux of AMS02~\cite{Vagelli:2016tph} is used to define the $\ep/e^\pm$ ratio upper bound from Eq.~(\ref{eq:posfromB}). On the right panel we show the upper bound on the $\ep$ flux.
We stress that Eq.~(\ref{eq:posfromB}) (and thus Fig.~\ref{fig:posfrac}), similarly to the $\pb/p$ situation exhibited in Fig.~\ref{fig:pb2}, is more sensitive to the unknown CR spectra in the spallation regions than is the $\ep/\pb$ ratio of Fig.~\ref{fig:pb2pos}.  
In Fig.~\ref{fig:pos2} we show how the bound is modified if one allows the proton spectrum in the secondary production sites to vary w.r.t. the locally measured proton flux. Models that realise this possibility include~\cite{Cowsik:2009ga,Burch:2010ye,Cowsik:2013woa,Blasi:2009hv,Blasi:2009bd,Mertsch:2009ph,Ahlers:2009ae,Kachelriess:2011qv,Cholis:2017qlb,Kachelriess:2015oua,Thomas:2016fcp}, reviewed in Sec.~\ref{ssec:secep}. The sensitivity of the bound in Fig.~\ref{fig:pos2} to proton flux variation should be compared to the insensitivity of the more robust $\ep/\pb$ ratio of Fig.~\ref{fig:pb2pos}.

\begin{figure}[t]
\begin{center}
\includegraphics[scale=0.6]{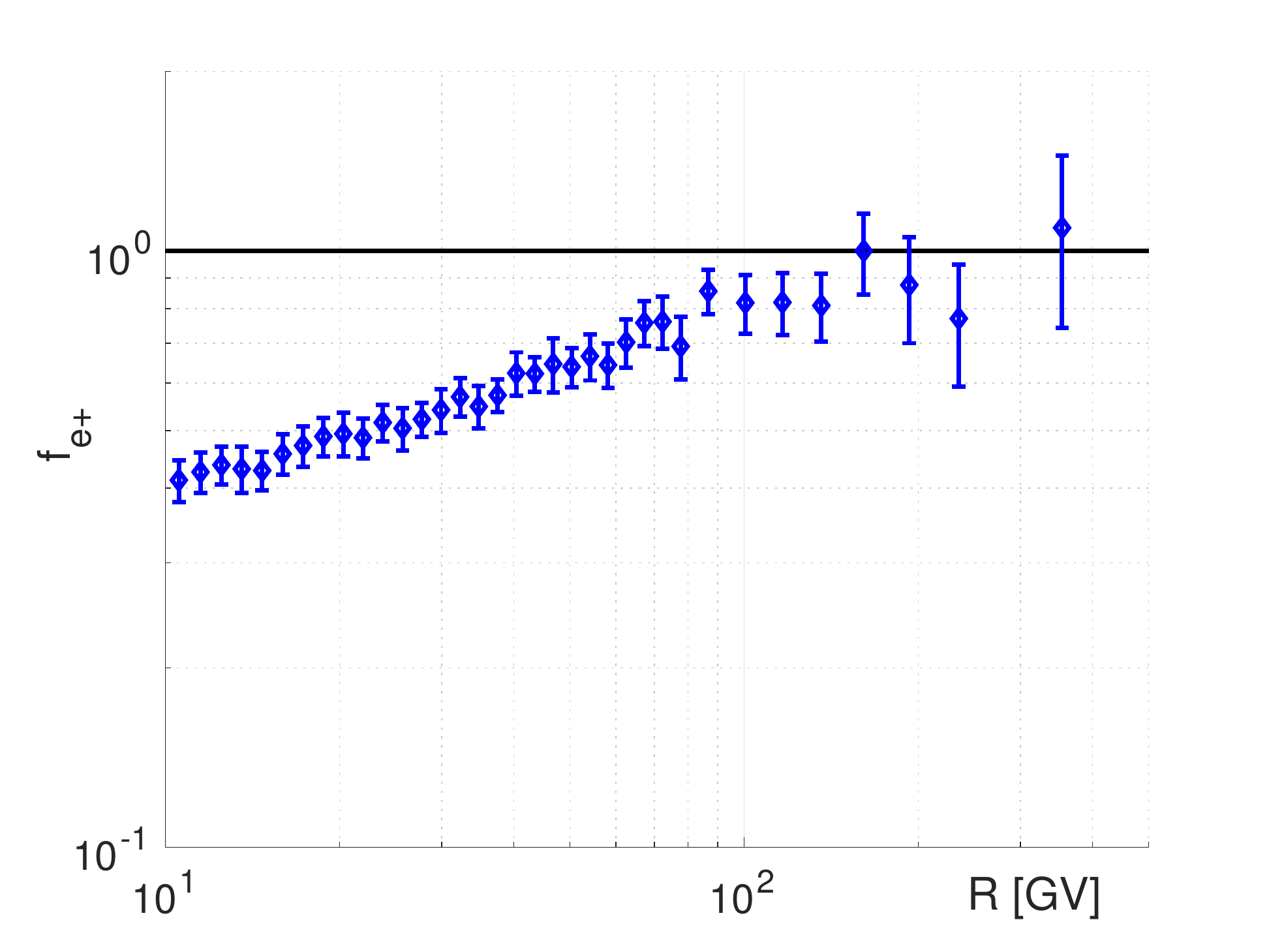}
\caption{$f_{\ep}$ extracted from Fig.~\ref{fig:pb2pos}. The upper bound for secondary $\ep$ reads $f_{\ep}\leq1$. Error bars reflect the measurement error on $\pb/\ep$ reported in~\cite{Aguilar:2016kjl}. Systematic cross section uncertainties in $pp\to\pb,\ep$, not shown in the plot, are in the ballpark of 10\%.}
\label{fig:fep}
\end{center}
\end{figure}

For later convenience it is useful to define the loss suppression factor $f_{\ep}$ via 
\be\label{eq:fep} \frac{n_{\ep}}{n_{\pb}}&=&f_{e^+}(\R)\frac{Q_{e^+}(\R)}{Q_{\bar p}(\R)}.\ee
In Fig.~\ref{fig:fep} we show $f_{\ep}$ as derived from Fig.~\ref{fig:pb2pos}. 
The upper bound for secondary $\ep$ is $f_{\ep}(\R)\leq1$.  
We find that $f_{\ep}(\R)$ is never much smaller than unity, and approaches unity from below for increasing $\R$. 
%
In considering this behaviour it is important to note that theoretically the possible range of the suppression factor is $0<f_{\ep}<\infty$: this just says that a prominent primary source of $e^{\pm}$ -- say, a nearby pulsar --  could make the $\ep$ flux as large as we wish in comparison to the  secondary upper bound; while strong radiative losses, if at work, could extinguish the flux. 

We conclude that AMS02 results hint for a secondary origin for $\ep$~\cite{Blum:2013zsa}\footnote{Ref.~\cite{Lipari:2016vqk} recently joined this understanding. We note, however, that our evaluation of the $Q_{\ep}/Q_{\pb}$ ratio in Fig.~\ref{fig:pb2pos} is higher than that of~\cite{Lipari:2016vqk} by 30-50\% at $\R\lesssim100$~GV. This difference led~\cite{Lipari:2016vqk} to conclude that $\ep$ are not affected by radiative losses at all energies; while we find that the data implies some radiative loss effect at $\R\lesssim100$~GV. The basic conclusion, putting 30-50\% differences aside, is similar: $\ep$ are consistent with secondaries.}. 
If $\ep$ are secondary, then Fig.~\ref{fig:pb2pos} suggests that the effect of energy loss in suppressing the $\ep$ flux is never very important, and possibly becomes less significant as we go to higher $\ep$ energy. As we shall see, this behaviour contradicts the expectations within common models of CR propagation\footnote{For early comprehensive analyses see e.g.~\cite{Protheroe:1982pp,Moskalenko:1997gh,Delahaye:2008ua}.}.
To appreciate this point we must venture into somewhat more muddy waters of CR astrophysics and consider the interplay of $\ep$ energy losses with the effects of propagation.

\subsection{Radiative energy loss vs. propagation time vs. grammage}\label{sec:tetc}
The name of the game is to figure out the interplay of $\ep$ energy loss with CR propagation: this is needed either to establish the necessity of a primary $\ep$ source, or to understand the lessons for CR propagation if $\ep$ are secondary.
 
The $e^+$ radiative cooling time is
\be\label{eq:tc}t_{\rm cool}(\R)&=&-\frac{\R}{\dot\R}\;\;.\ee
At high energy $\R>10$~GV, energy loss is dominated by synchrotron and IC. In the Thomson regime~\cite{Blumenthal:1970gc}
\be\label{eq:tcThom}\tc(\R)&=&\frac{3m_e^4}{4\sigma_TU_T}\approx30\left(\frac{\R}{10~{\rm GV}}\right)^{-1}\left(\frac{U_T}{1~{\rm eVcm^{-3}}}\right)^{-1}~{\rm Myr},\ee
where $U_T$ is the sum of radiation and magnetic field energy densities. Thus, in the Thomson regime $t_{\rm cool}\propto\R^{-1}$. 
Bremstrahlung and Klein-Nishina corrections soften this behaviour to $t_{\rm cool}\sim\R^{-\gamma_c}$ with $\gamma_c<1$, as we discuss in Sec.~\ref{ssec:secepprop}.\\

Consider the qualitative behaviour of $f_{e^+}$. We expect $f_{\ep}$ to increase monotonically as a function of $t_{\rm cool}/t_{\rm esc}$, where $t_{\rm esc}$ is the CR propagation time, defined in some convenient way to parametrise the typical time a CR spends in the system from the time of production until the time of detection at Earth. 

In the limit $t_{\rm cool}/t_{\rm esc}\gg1$, we expect $f_{e^+}\to1$; this is because in this limit, a typical CR trajectory lasts much less time than the time it takes an $\ep$ to lose a significant amount of its initial energy. Thus in this limit relativistic $\ep$ and $\pb$ at the same $\R$ propagate in the same way and the observed $\ep/\pb$ ratio reflects the secondary production rate ratio $Q_{\ep}/Q_{\pb}$. In the opposite limit, $t_{\rm cool}/t_{\rm esc}\ll1$, we expect $f_{e^+}\ll1$ because $\ep$ lose most of their energy on their way from production to detection, while the corresponding $\pb$ propagate unaffected.  \\

Given an estimate of $t_{\rm cool}$, Fig.~\ref{fig:fep} is therefore a measurement of the CR propagation time. 
The detailed interpretation of the form of $f_{e^+}$, however, is model-dependent. 
To demonstrate this point, in App.~\ref{app:demo} we calculate $f_{e^+}$ for two propagation model examples: a version of the leaky-box model (LBM), and a one-dimensional thin disc+halo diffusion model. 

\begin{figure}[!h]\begin{center}
\includegraphics[width=0.8\textwidth]{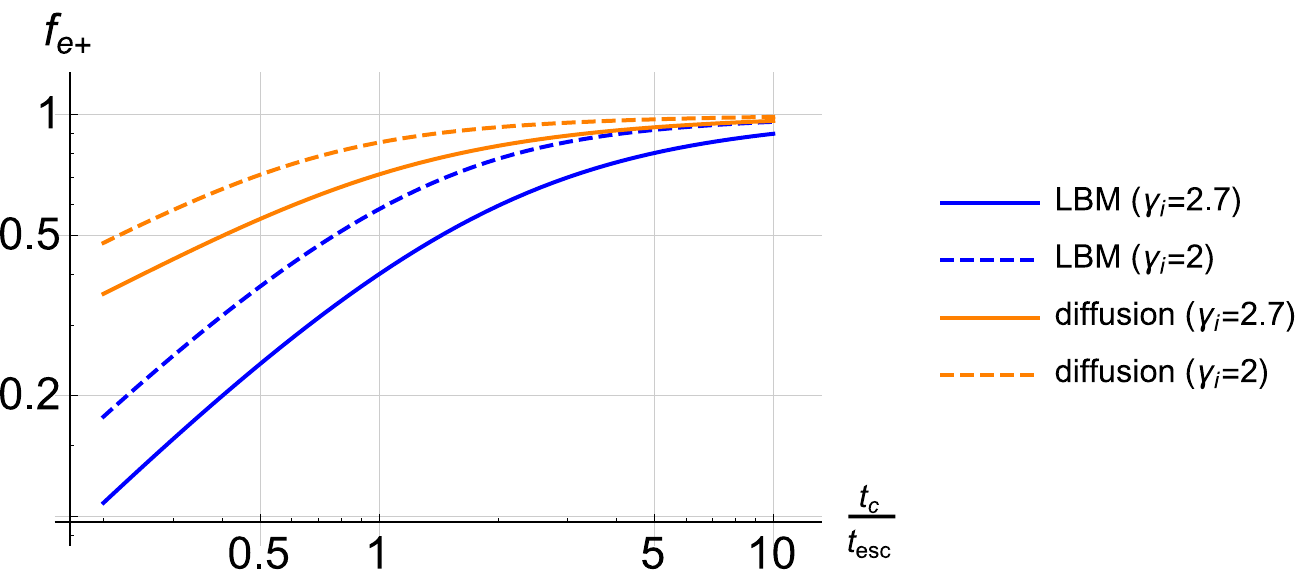}
\end{center}
\caption{The $e^+$ flux loss suppression factor $f_{e^+}$, as function of the cooling to escape time ratio, for the LBM and diffusion models with different values of the primary proton spectral index $\gamma_i$ in the secondary production region. Details of the calculation are given in App.~\ref{app:demo}.}
\label{fig:fpos_tcte}
\end{figure}

We emphasize that both of these models satisfy Eq.~(\ref{eq:sec}). Therefore, calibrating the relevant free parameters in either model to reproduce $\X\sim\R^{-0.4}$ according to Fig.~\ref{fig:X} would make these models automatically reproduce Fig.~\ref{fig:pb}, consistent with AMS02 $\pb/p$ data. 
Fig.~\ref{fig:fpos_tcte} shows $f_{\ep}$ as obtained for the two models, with propagation parameters calibrated consistently with B/C and $\pb/p$. 
We take representative values of $\gamma_i=2$ and 2.7 for the primary proton spectral index. The numerical value of $f_{\ep}$ differs between the LBM and diffusion models, despite having calibrated both of these models to match B/C and $\pb/p$. 
%

%

While the diffusion model and the LBM differ in the form of $f_{e^+}$ they predict,  both models share a common feature: 
in both of the models, the rigidity-dependent column density of ISM traversed by CRs is proportional to the rigidity-dependent propagation time, $\X(\R)\propto\te(\R)$.

This proportionality between $\Xe$ and $\te$ is more general than the specific models we looked at. 
It holds, for example, for commonly adopted diffusion models~\cite{Strong:2007nh} that assume a rigidity independent diffusion boundary. Because the grammage $\Xe$ must be fit in phenomenologically consistent versions of these models\footnote{By adjusting the free parameters $K(\R)$, $L$, and other parameters in more complicated realisations. E.g., the inhomogeneous diffusion model of~\cite{Kappl:2016qug} falls in the same category, and is thus affected by the same problem in reconciling $\ep$ and B or $\pb$ data: it is constrained by construction to satisfy Eq.~(\ref{eq:Xevste}).} to match B/C, sub-Fe/Fe, and $\pb/p$ data, these models satisfy
\be\label{eq:Xevste}\te(\R)\propto\Xe(\R)\propto\R^{-0.4},\ee
where in the numerical assignment we adopt, for simplicity, the approximate fit of Ref.~\cite{Blum:2013zsa}.

Using Eq.~(\ref{eq:tcThom}) gives $\tc(\R)\propto\R^{-\gamma_c}$ with $\gamma_c=1$, so propagation models that satisfy Eq.~(\ref{eq:Xevste}) predict that the ratio $t_{\rm cool}(\R)/\te(\R)\propto\R^{0.4-\gamma_c}\sim\R^{-0.6}$ must decrease with increasing rigidity. As a result, because $f_{\ep}$ scales as a positive power of $t_{\rm cool}(\R)/\te(\R)$, these models predict that the effect of losses should become increasingly more important at high energy: $f_{\ep}(\R)$ should {\it decrease} at rising $\R$. This is the opposite trend to that inferred from data in Figs.~\ref{fig:pb2pos}-\ref{fig:fep}.\\

To maintain the hypothesis of secondary $\ep$, the fact that the observed $f_{\ep}$ approaches unity with increasing rigidity implies that the propagation time $\te$ cannot be much larger than the cooling time $t_{\rm cool}$, and is decreasing with rigidity faster than $\tc$. This means that {\it$\te$ decreases with rigidity faster than} $\Xe$, in contradiction to Eq.~(\ref{eq:Xevste}).\\

We are faced with two possibilities. Either take the coincidence of the observed $\ep/\pb$ ratio with pp branching fractions (Figs.~\ref{fig:pb2pos}-\ref{fig:fep}) as a hint that CR $\ep$ are secondary, in which case something basic is not right in the commonly adopted CR diffusion models: the CR grammage cannot be proportional to the CR propagation time. Or, accept the diffusion models and invoke a primary source for $\ep$, like dark matter annihilation or pulsars, in which case some primary source parameters would need to be tuned to reproduce the observed $\ep/\pb$ ratio as an accident.

This is a good point to comment on statements in the literature, that advocated the presence of an $\ep$ primary source based on a rising $\ep/e^{\pm}$ fraction. Two representative examples are~\cite{Adriani:2008zr} and~\cite{Serpico:2008te}. Ref.~\cite{Adriani:2008zr}, and numerous following works, based the call for primary $\ep$ on the observation that a rising $\ep/e^\pm$ would be inconsistent with secondary $\ep$ as expected in a certain diffusion model~\cite{Moskalenko:1997gh}. However, phenomenological models such as~\cite{Moskalenko:1997gh} are constructed with many simplifying assumptions, ranging from steady state and homogeneous diffusion to the geometry and boundary conditions of the CR halo~\cite{Strong:2007nh}. Some of these assumptions may not apply to Nature. We  explore alternative ideas with secondary $\ep$ in Sec.~\ref{ssec:secep}. 

Ref.~\cite{Serpico:2008te} argued that a rising $\ep/e^\pm$ requires primary $\ep$ because $e^-$ and $\ep$ suffer radiative energy losses in the same way, and because the primary $e^-$ source spectrum cannot be softer than that of the secondary $\ep$. The problem with this argument is that the injection spectrum of primary $e^-$ is unknown\footnote{Models like~\cite{Moskalenko:1997gh}, for example, take the $e^-$ injection spectrum as {\it a free parameter that is then fitted to the $e^-$ data} (similar practice -- with separate free parameters -- is applied for the proton, He, and other primary nuclei spectra).}, and the propagation paths of $\ep$ and $e^-$ may differ because their production sites as secondaries vs. accelerated primaries, respectively, may be different. In this case, $e^-$ may suffer additional energy losses as compared to $\ep$. We will see model examples for this, too, in Sec.~\ref{ssec:secep}.\\

Finally, it is important to note that the high energy $\ep$ data~\cite{Adriani:2008zr,PhysRevLett.110.141102,AMS02C2O:2016} represent a new observational probe of CR propagation at high rigidity $\R>10$~GV: it tests the propagation models where they were not tested before.   
B/C data measures $\Xe$; as we have seen, other stable secondary nuclei data such as $\pb/p$ do not give much of a new test: they are essentially consistency checks of the hypothesis that different CR species sample similar $\Xe$, a fact for which early evidence already existed~\cite{Engelmann:1990zz,2003ApJ...599..582W,1992ApJ...394..174G}. In contrast, the $\ep$ data is sensitive to the a-priori independent quantity $\te$. The fact that $\ep$ could eventually provide such a test of the models was pointed out already in~\cite{Ginzburg:1976dj,Ginzburg:1990sk}, long before PAMELA and AMS02 made this test come to life.

Besides from $\ep$, no other CR data to date accesses $\te$ in the same range in rigidity $\R\gsim100$~GV. What comes nearest are measurements of the effect of radioactive decay of secondary Be, Al, and Mg isotopes. 
In Sec.~\ref{sec:radnuc} we consider these data as a model-independent test of the secondary $\ep$ hypothesis. As of today, the test is based on early measurements~\cite{1998ApJ...506..335W,Engelmann:1990zz} and supports the idea of secondary $\ep$~\cite{Katz:2009yd}, but is inconclusive due to systematic uncertainties. A better test should become possible in the near future with AMS02 data, and we devote some time to explain the key physics.

\subsection{A test with radioactive nuclei}\label{sec:radnuc}
Measurements of the suppression of the flux of secondary radioactive nuclei due to decay in flight constrain the  CR propagation time $\te$~\cite{1970ARNPS..20..323S,1998ApJ...506..335W,1999ICRC....4..195P,1988SSRv...46..205S,1998AA...337..859P,2001ApJ...563..768Y,Donato:2001eq,Putze:2010zn,Blum:2010nx,2007ApJ...655..892S}. 
%
The idea is that a relativistic radioactive nucleus with rest frame lifetime $\tau_d$, mass number $A$ and charge $eZ$ has an observer frame lifetime 
\be\label{eq:td} t_d(\R)&\approx&\sqrt{1+\frac{e^2Z^2\R^2}{A^2m_p^2}}\tau_d,\ee
when observed as CR. Given a secondary radioactive species like $^{10}$Be, we can compute the prediction for its density, discarding the effect of decay, and compare the result to the observed density. This allows to define a decay suppression factor that can be extracted directly from data:
\be\label{eq:fbe10} f_{^{10}\rm Be}(\R)&=&\frac{n_{\rm ^{10}Be}(\R)}{n_{\rm ^{10}Be,\,no\,decay}(\R)}\approx\frac{n_{\rm ^{10}Be}(\R)}{\frac{Q^+_{\rm ^{10}Be}(\R)}{1+\Xe(\R)\frac{\sigma_{^{10}\rm Be}}{m}}}.\ee
Here, the numerator is supposed to be taken directly from measurement, while the denominator is a theory output but is, again, data-driven based on fragmentation cross sections, B/C and primary CR spectra. 
In the last expression we used Eqs.~(\ref{eq:sec}) and~(\ref{eq:Xbc}), and noted that for stable secondary species $i$ Eq.~(\ref{eq:sec}) implies
\be n_i(\R)&\approx&\frac{Q^+_i(\R)}{1+\Xe(\R)\frac{\sigma_i}{m}},\ee
where $Q^+_i(\R)$ is the positive production term in the net source $Q_i$. (As before, $\sigma_i/m$ is the total inelastic cross section of $i$ per ISM particle mass.)\\

We expect the qualitative behaviour of $f_{^{10}\rm Be}(\R)$ to depend on the CR propagation time $\te$ via the ratio $t_d(\R)/\te(\R)$: for $t_d(\R)/\te(\R)\to\infty$ we should have $f_{^{10}\rm Be}(\R)\to1$, while for $t_d(\R)/\te(\R)\to0$ we expect $f_{^{10}\rm Be}(\R)\ll1$. Because $\te(\R)$ decreases with increasing $\R$, and $t_d(\R)$ (for relativistic nuclei) increases as $t_d\propto\R$, it is clear that $f_{^{10}\rm Be}(\R)$ should approach unity with increasing $\R$. Since $t_d$ is known from laboratory data, a measurement of $f_{^{10}\rm Be}$ provides a constraint on $\te$. 

Experimental data on radioactive nuclei is divided in two categories: isotopic data such as $^{10}$Be/$^{9}$Be, and elemental (or charge) data like Be/B in which the numerator denotes the sum of Be isotopes ($^{7,9,10}$Be) and the denominator the sum of B isotopes ($^{10,11}$B). 

Isotopic ratios are experimentally challenging to measure at high energy. As a result, current data on $^{10}$Be/$^{9}$Be is limited to low rigidity $\R\lesssim1$~GV. This introduces significant theory uncertainty in the interpretation of these data as effects such as solar modulation, energy-dependence in the nuclear fragmentation cross sections, and various propagation effects that change CR energy during propagation become important (see e.g.~\cite{1996ApJ...465..972P}). Nevertheless, diffusion models have traditionally used this low energy data to calibrate the models~\cite{Strong:2007nh}, extrapolating the result to the relativistic range. AMS02 is expected to improve the situation with the ability to measure $^{10}$Be/$^{9}$Be up to $\R\sim10$~GV.

Elemental ratios can be measured to high energy~\cite{1998ApJ...506..335W}. The challenge here is that the contribution of the radioactive $^{10}$Be isotope to the total Be flux is never expected to exceed about $\sim30\%$, based on our knowledge of partial fragmentation cross sections like $^{12}$C$\to^{10}$Be vs. $^{12}$C$\to^{7,9}$Be, etc. As a result, the Be/B ratio is limited to the range $0.3\lesssim{\rm Be/B}\lesssim0.4$ or so, making the interpretation particularly sensitive to fragmentation cross section uncertainties. The situation with other relevant elemental ratios, Al/Mg and Cl/Ar, is similar\footnote{For the Al and Cl, another difficulty is that primary contamination to the flux may not be negligible~\cite{Blum:2010nx}.}.  

\begin{figure}[!h]\begin{center}
\includegraphics[width=0.65\textwidth]{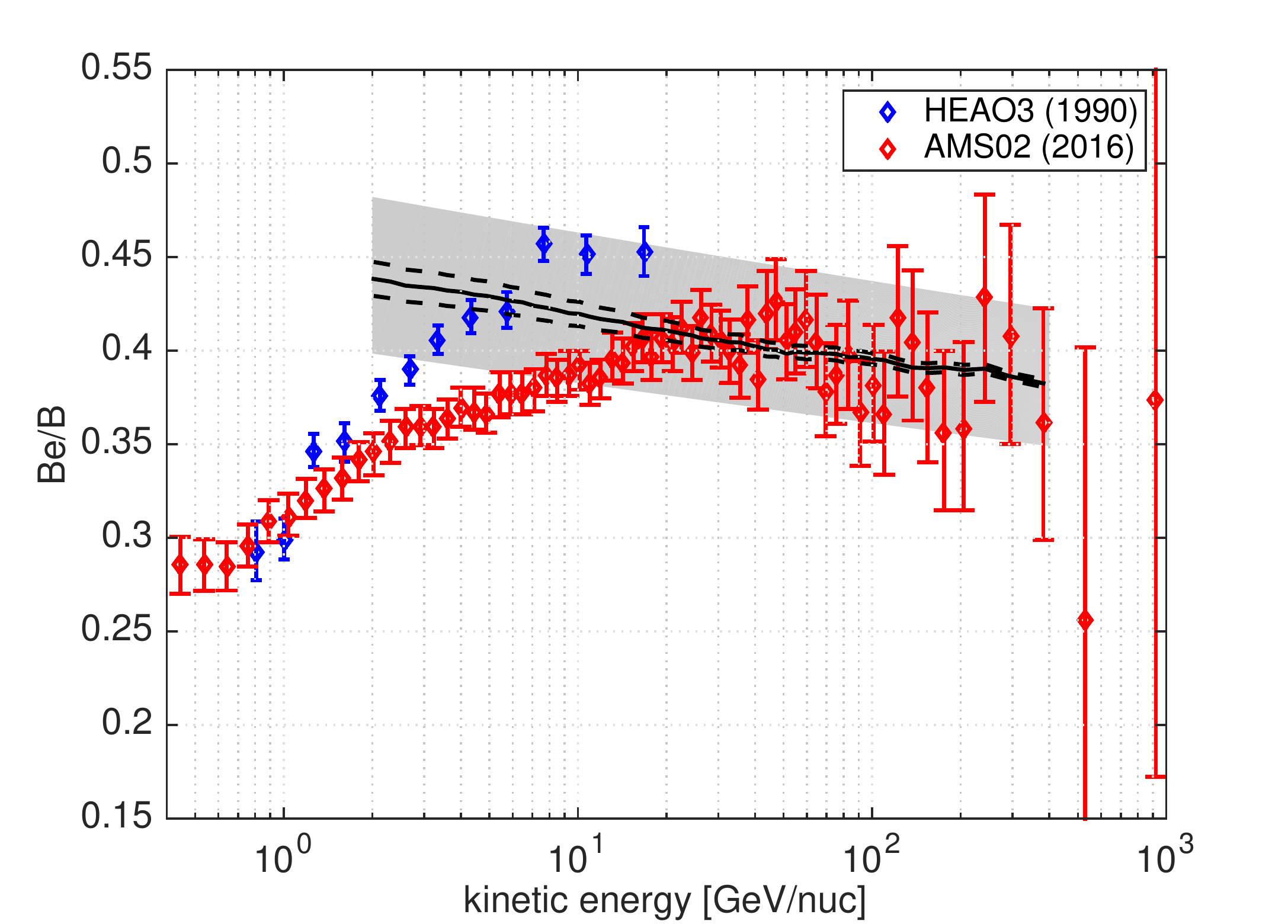}
\caption{The elemental flux ratio Be/B. HEAO3 data shown in blue; preliminary data from AMS02 in red. The no-loss secondary prediction is shown by black line. Shaded band shows the effect of varying by $\pm10$\% the cumulative cross section for B production, $\sigma_{\rm CNO\to B}$. Dashed lines show the effect of varying the cross sections for $^{11}$B$\to^{7,9,10}$Be, in a correlated way, by $\pm20$\%.}
\label{fig:Be2Binf}
\end{center}
\end{figure}

The Be/B ratio derived from early measurements of Be and B fluxes by the HEAO3 mission is shown in Fig.~\ref{fig:Be2Binf} (blue markers)~\cite{1998ApJ...506..335W,Engelmann:1990zz}. Recently, AMS02 reported preliminary results for Be/B~\cite{AMS02C2O:2016} extending to very high energy, shown in red. The saturation point $f_{^{10}\rm Be}\to1$ is, for the first time, clearly manifest in this data as the observed Be/B saturates the no-decay secondary prediction\footnote{It is interesting to note that the theoretically predicted asymptotic no-decay Be/B ratio, shown by the grey band in Fig.~\ref{fig:Be2Binf}, does not go to a constant at high $\R$ but rather exhibits a mild decrease with increasing $\R$. This trend is caused by the so-called tertiary production where $^{11}$B fragments into $^{7,9,10}$Be; this effect is contained in Eq.~(\ref{eq:sec}), and the mild decrease in asymptotic Be/B is due to the decrease of $\Xe$ at rising $\R$. The preliminary AMS02 data is consistent with this subtle prediction of Eq.~(\ref{eq:sec}). We await an official release by AMS02 for further analysis.}.

\begin{figure}[!h]\begin{center}
\includegraphics[width=0.6\textwidth]{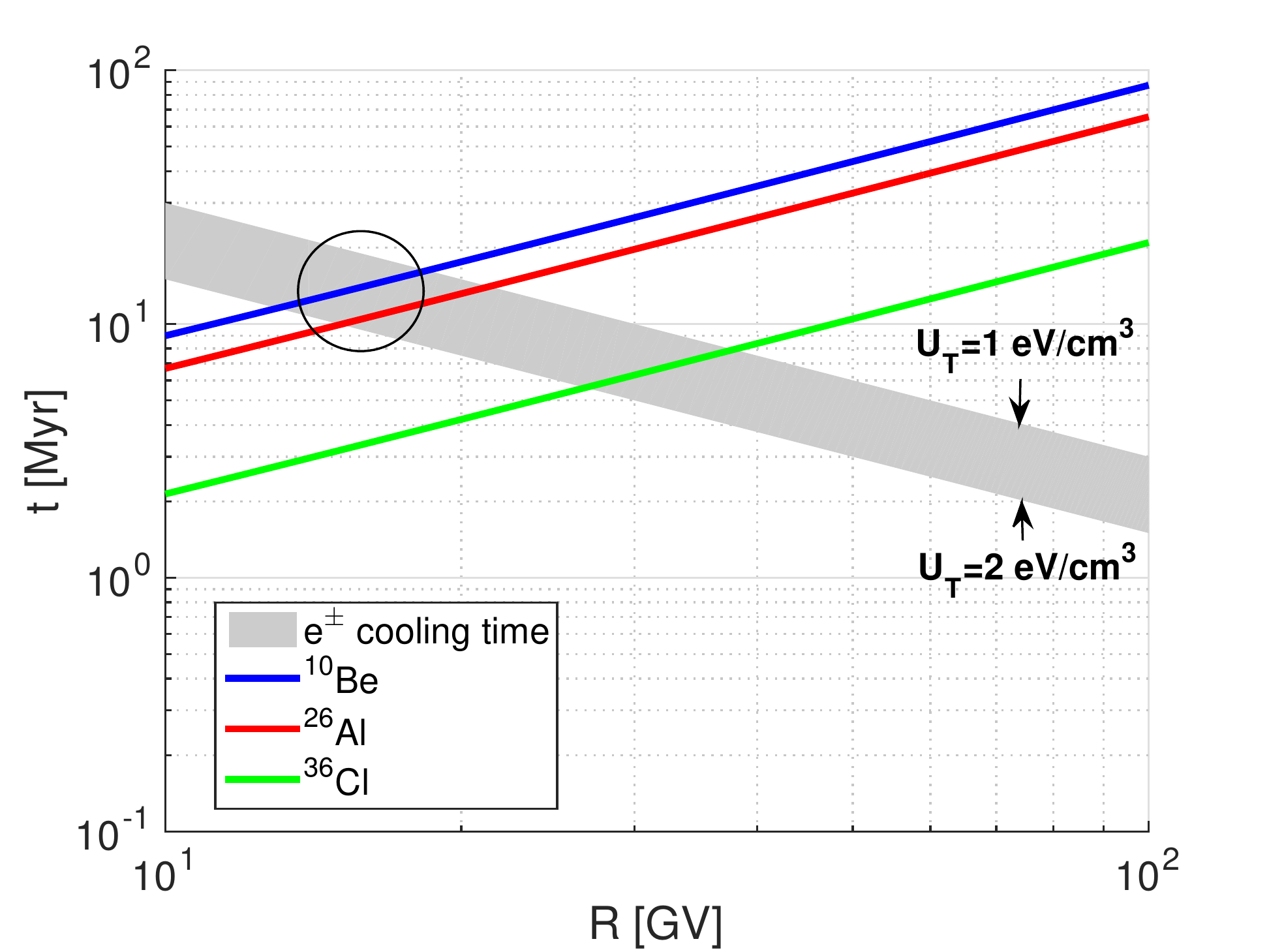}
\caption{Radioactive decay vs. $\ep$ energy loss. An estimate of the $\ep$ cooling time $\tc$ is shown by grey band, obtained for $U_T$ between 1-2 eV/cm$^3$, neglecting Klein-Nishina corrections. The observer frame lifetimes of CR $^{10}$Be, $^{26}$Al, and $^{36}$Cl are shown by blue, red, and green lines, respectively. Around $\R\sim15$~GV (kinetic energy per nucleon $k\sim5$~GeV/nuc), the observer frame lifetime of $^{10}$Be approximately coincides with the cooling time of CR $\ep$.}
\label{fig:tctddemo}
\end{center}
\end{figure}

The definition of the decay suppression factor $f_{^{10}\rm Be}(\R)$ is analogous to that of the loss suppression factor $f_{\ep}(\R)$ defined in Eq.~(\ref{eq:fep}) for secondary $\ep$. However, radioactive decay is not quite the same as radiative energy loss: the former eliminates the CR altogether, while the latter just degrades it in energy. Moreover, the $\R$ dependence of $t_d(\R)\sim\R$ for radionuclei is essentially opposite to $\tc(\R)\sim\R^{-1}$ for $\ep$. 
Fig.~\ref{fig:tctddemo} illustrates the behaviour of $t_d$ and $\tc$.

Ref.~\cite{Katz:2009yd} pointed out an approximate, but model independent way in which radiative energy losses and radioactive decay can be put on similar footing. Consider the contribution of radioactive decay in a general, local transport equation. Decay introduces a term to the continuity equation for radionucleus $i$,
\be\left(\frac{\partial n_i}{\partial_t}\right)_{\rm decay}&=&-\frac{n_i(\R)}{t_{d,i}(\R)}.\ee
Energy loss for $\ep$, in comparison, is captured by:
\be\left(\frac{\partial n_{\ep}}{\partial_t}\right)_{\rm energy\,loss}&=&\frac{\partial}{\partial\R}\left(\dot\R n_{\ep}\right)=-\frac{n_{\ep}(\R)}{\tilde t_{\rm cool}(\R)},\ee
where we define
\be\label{eq:ttc}\frac{1}{\tilde t_{\rm cool}(\R)}&=&-\left[\frac{\partial\log\left(\frac{\R n_{\ep}}{\tc}\right)}{\partial\log\R}\right]\frac{1}{\tc(\R)}.\ee
The observed $\ep$ flux in the range $\R\sim(10-500)$~GV is roughly a power law $n_{\ep}\sim\R^{-\gamma_{\ep}}$ with $\gamma_{\ep}$ in the range $(2.75-3)$. For $\tc$ not far from the Thomson regime, $\tc\sim\R^{-1}$, the logarithmic term in Eq.~(\ref{eq:ttc}) is therefore a weak function of $\R$, varying in the range $-\left[\frac{\partial\log\left(\R^2n_{\ep}\right)}{\partial\log\R}\right]\sim(0.75-1)$. 
We learn that, due to the steeply falling $\ep$ flux, decay and energy loss are represented by a similar form in the continuity equation. \\

A model independent check of the hypothesis of secondary $\ep$ is therefore obtained by comparing the observed $f_{\ep}(\R^*)$ for $\ep$, and $f_{i}(\R^*)$ for radionucleus $i$, at the particular rigidity $\R=\R^*$ in which $t_{d,i}(\R^*)=\tc(\R^*)$. 
Referring to Fig.~\ref{fig:tctddemo}, we see that for $^{10}$Be, with a reasonable estimate of $U_T$, $\R^*\sim15$~GV. This is illustrated by a circle in the plot. For this rigidity, Fig.~\ref{fig:fep} suggests 
\be f_{\ep}(15~{\rm GV})&\sim&0.4-0.5.\ee
From the HEAO3 data~\cite{Engelmann:1990zz} analysis of~\cite{1998ApJ...506..335W}, Ref.~\cite{Katz:2009yd} found
\be f_{^{10}\rm Be}(15~{\rm GV})&\sim&0.3-0.4.\ee
Consistent results obtain for the Al/Mg and Cl/Ar data. 

Significant systematic uncertainty affects this analysis, manifest in Fig.~\ref{fig:Be2Binf} by the cross section uncertainty as well as the mismatch between HEAO3 and AMS02 preliminary results. Nevertheless, we can conclude that this test of radionuclei data is consistent with secondary $\ep$. Upcoming results from AMS02~\cite{AMS02C2O:2016} should allow to improve this test.

Finally, Ref.~\cite{Blum:2010nx} suggested that the rigidity dependence of $\te$ can be constrained by comparing the decay suppression factor $f_i$ for different species of radionuclei measured at the same observer frame lifetime and thus -- because different nuclei have different rest frame lifetimes -- at different rigidities.

\begin{figure}[!h]\begin{center}
\includegraphics[width=0.475\textwidth]{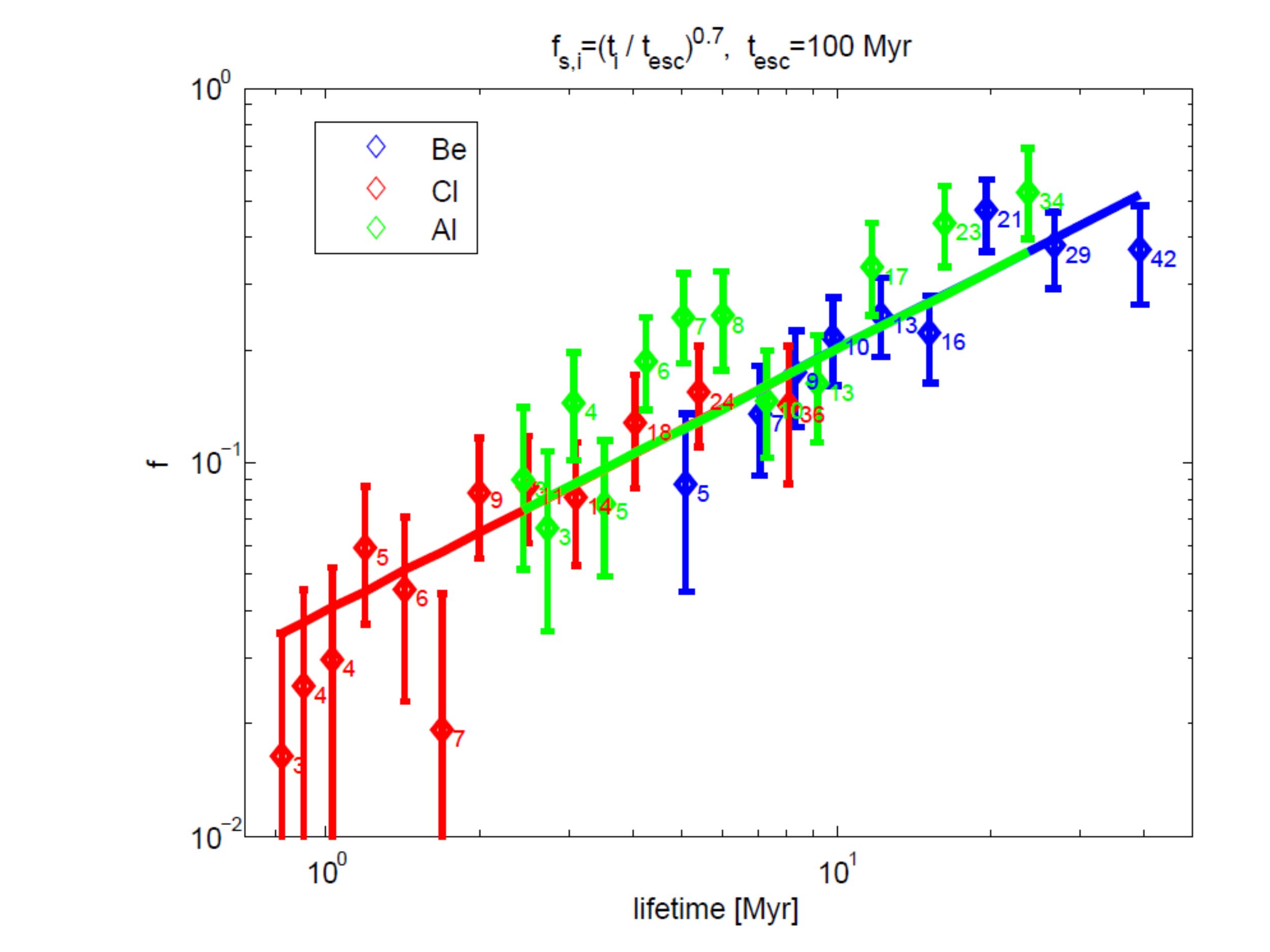}\quad
\includegraphics[width=0.475\textwidth]{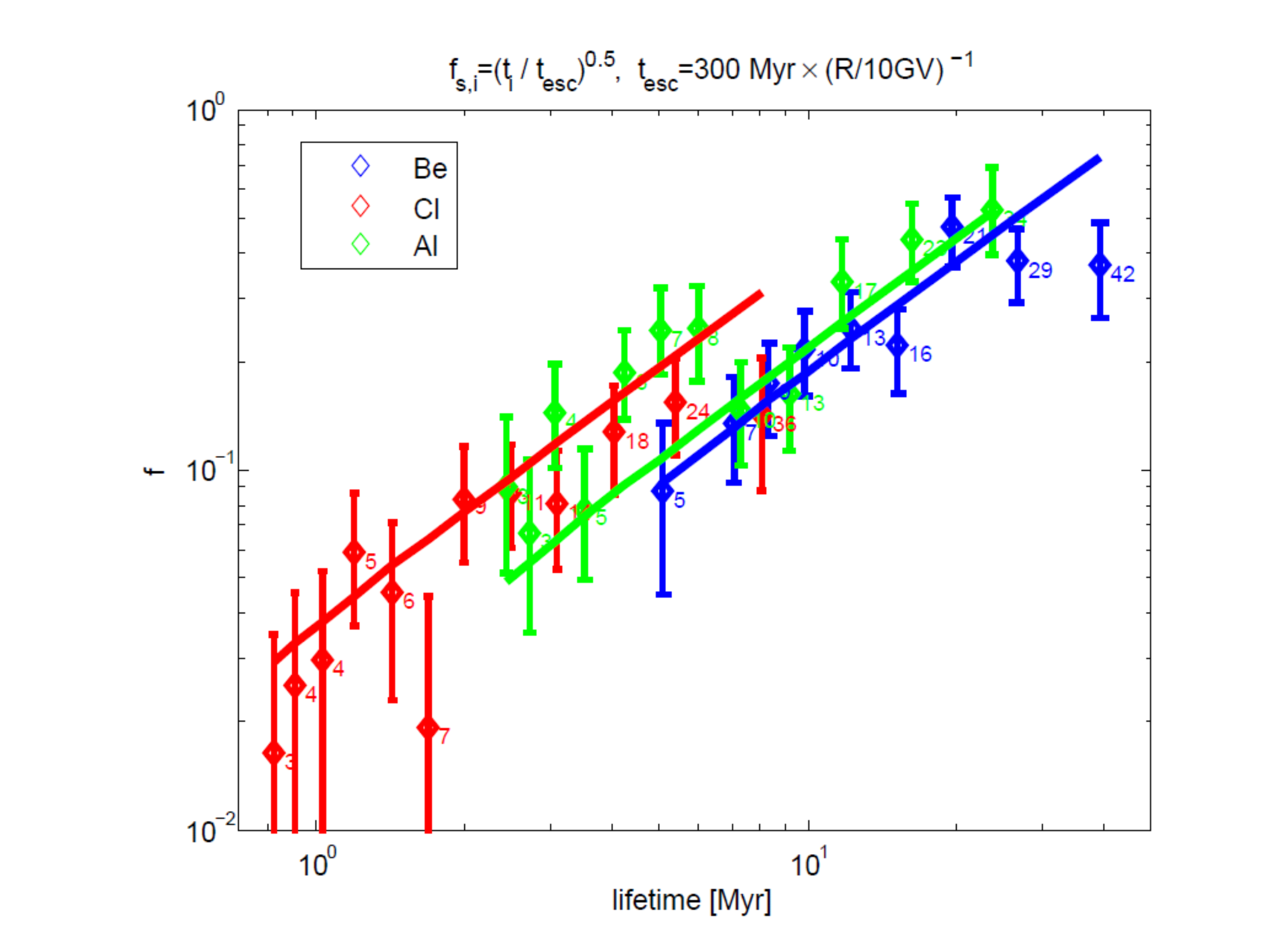}
\caption{Radioactive nuclei data from HEAO3~\cite{Engelmann:1990zz,1998ApJ...506..335W}, presented in terms of the decay suppression factor $f_i$ vs observer frame lifetime derived from Eq.~(\ref{eq:td}). Numbers next to each point denote the rigidity $\R$ (in GV) for that nucleus species at that observer frame lifetime. {\bf Left:} fit assuming $\R$-independent $\te$. If this turns out to be the correct fit, {\it then CR $\ep$ are not secondary}. {\bf Right:} fit assuming $\te\propto\R^{-1}$, roughly as needed to obtain $f_{\ep}$ that is rising or flat with $\R$. Current uncertainties do not allow a clear preference, but upcoming AMS02 data, with some improvement in fragmentation cross section analyses (see e.g.~\cite{Tomassetti:2015nha}), can change the picture. From Ref.~\cite{Blum:2010nx}.}
\label{fig:timesNuc}
\end{center}
\end{figure}

Fig.~\ref{fig:timesNuc} illustrates this point with HEAO3 data~\cite{Engelmann:1990zz,1998ApJ...506..335W}. While, again, systematic and statistical uncertainties in this data are large, the idea is promising: a rapidly decreasing $\te(\R)$, as needed to reconcile secondary $\ep$ with a rising $f_{\ep}(\R)$, would manifest in the radionuclei data by resolving the combined radionuclei data set into three different $f_i$ curves for Be/B (controlled by $\tau_{^{10}\rm Be}=2.2$~Myr), Al/Mg ($\tau_{^{26}\rm Al}=1.3$~Myr), and Cl/Ar ($\tau_{^{36}\rm Cl}=0.44$~Myr). 

\subsection{Implications of secondary $\ep$ for CR propagation}\label{ssec:secepprop}
In this section we assume that $\ep$ are secondary, and review constraints on CR propagation that can be deduced in this case.
Fig.~\ref{fig:fep} suggests that 
\be\label{eq:fep300} f_{\ep}(\R=300~{\rm GV})\gsim0.7,\\
\label{eq:fep10} f_{\ep}(\R=10~{\rm GV})\lsim0.5.\ee 
Considering the two models depicted in Fig.~\ref{fig:fpos_tcte} as representative examples, Eq.~(\ref{eq:fep300}) implies 
$\tc(\R=300~{\rm GV})/\te(\R=300~{\rm GV})\gsim1$ for the diffusion model, and $\tc(\R=300~{\rm GV})/\te(\R=300~{\rm GV})\gsim3$ for the LBM. 
Eq.~(\ref{eq:fep10}) implies 
$\tc(\R=10~{\rm GV})/\te(\R=10~{\rm GV})\lsim0.4$ for the diffusion model, and $\tc(\R=10~{\rm GV})/\te(\R=10~{\rm GV})\lsim1.5$ for the LBM.

Ignoring Klein-Nishina corrections (but see discussion below), we summarise these results by the  constraints~\cite{Blum:2013zsa}\footnote{Ref.~\cite{Lipari:2016vqk} recently arrived at comparable conclusions.},
\be\label{eq:tetc} t_{\rm esc}\left(\R=300~{\rm GV}\right)&\le &t_{\rm cool}\left(\R=300~{\rm GeV}\right)\no\\
\label{eq:tetc200}&\sim&1{\rm~Myr}\left(\frac{\bar U_T}{{\rm 1~eVcm^{-3}}}\right)^{-1},\\
t_{\rm esc}\left(\R=10~{\rm GV}\right)&>&t_{\rm cool}\left(\R=10~{\rm GeV}\right)\no\\
\label{eq:tetc10}&\sim&30{\rm~Myr}\left(\frac{\bar U_T}{{\rm 1~eVcm^{-3}}}\right)^{-1}.\ee
The RHS of Eqs.~(\ref{eq:tetc200}-\ref{eq:tetc10}) is based on a rough estimate of the $e^\pm$ cooling time at the relevant energies, and as such is subject to O(1) uncertainty. 
Here $\bar U_T$ is the time-averaged total electromagnetic energy density (propagation path- and time-average of $U_T$ from Eq.~(\ref{eq:tcThom})) in the propagation region. It is natural to expect that $\bar U_T$ depends on CR rigidity, both because the radiation and magnetic fields in the ISM are not uniformly distributed and because the Thomson limit for describing the losses is not exact. 

One irreducible source for energy dependence in the effective value of $\bar U_T$ comes from Klein-Nishina corrections, that are neglected in Eqs.~(\ref{eq:tetc200}) and~(\ref{eq:tetc10}). 
The Thomson limit is not a good approximation for 20-300~GV positrons if $U_T$ contains a significant UV component, as may be the case judging from estimates of the Milky Way radiation field~\cite{Hauser:2001xs,Porter:2005qx,2010ApJ...725..466C}. In terms of Eqs.~(\ref{eq:tetc200}-\ref{eq:tetc10}), a plausible $\sim$50\% UV contribution to $U_T$ implies that the effective value of $\bar U_T$ decreases between 10~GV to 300~GV by a factor of $\sim$2. More extreme possibilities were entertained in~\cite{Stawarz:2009ig}.

Another feature that is not included in Eqs.~(\ref{eq:tcThom},\ref{eq:tetc200}-\ref{eq:tetc10}) is bremsstrahlung losses. The bremsstrahlung radiation length is $\zeta\approx60$~g/cm$^2$~\cite{Longair:1992ze}, approximately independent of energy and insensitive to the H:He ratio in the ISM, such that the corresponding cooling time is $t_{\rm brem}\approx\zeta/(c\rho_{ISM})$. The energy loss term in the $\ep$ continuity equation takes a form similar to that of a fragmentation loss term for nuclei,
\be\label{eq:brem}\frac{\partial}{\partial\R}\left(\dot\R n_\ep\right)_{\rm brem}&\approx&
-\Gamma_{\rm brem}\,n_\ep,\ee
where 
\be\label{eq:Gbrem}\Gamma_{\rm brem}&=&-\frac{c\rho_{ISM}}{\zeta}\left[\frac{\partial\log\left(\R n_\ep\right)}{\partial\log\R}\right].\ee
For $n_{\ep}\sim\R^{-\gamma_{\ep}}$ with $\gamma_{\ep}\sim2.75-3$, we have $\Gamma_{\rm brem}\approx C(c\rho_{ISM}/\zeta)$ with $C\sim1.75-2$. 

\begin{figure}[t]
\begin{center}
\includegraphics[scale=0.6]{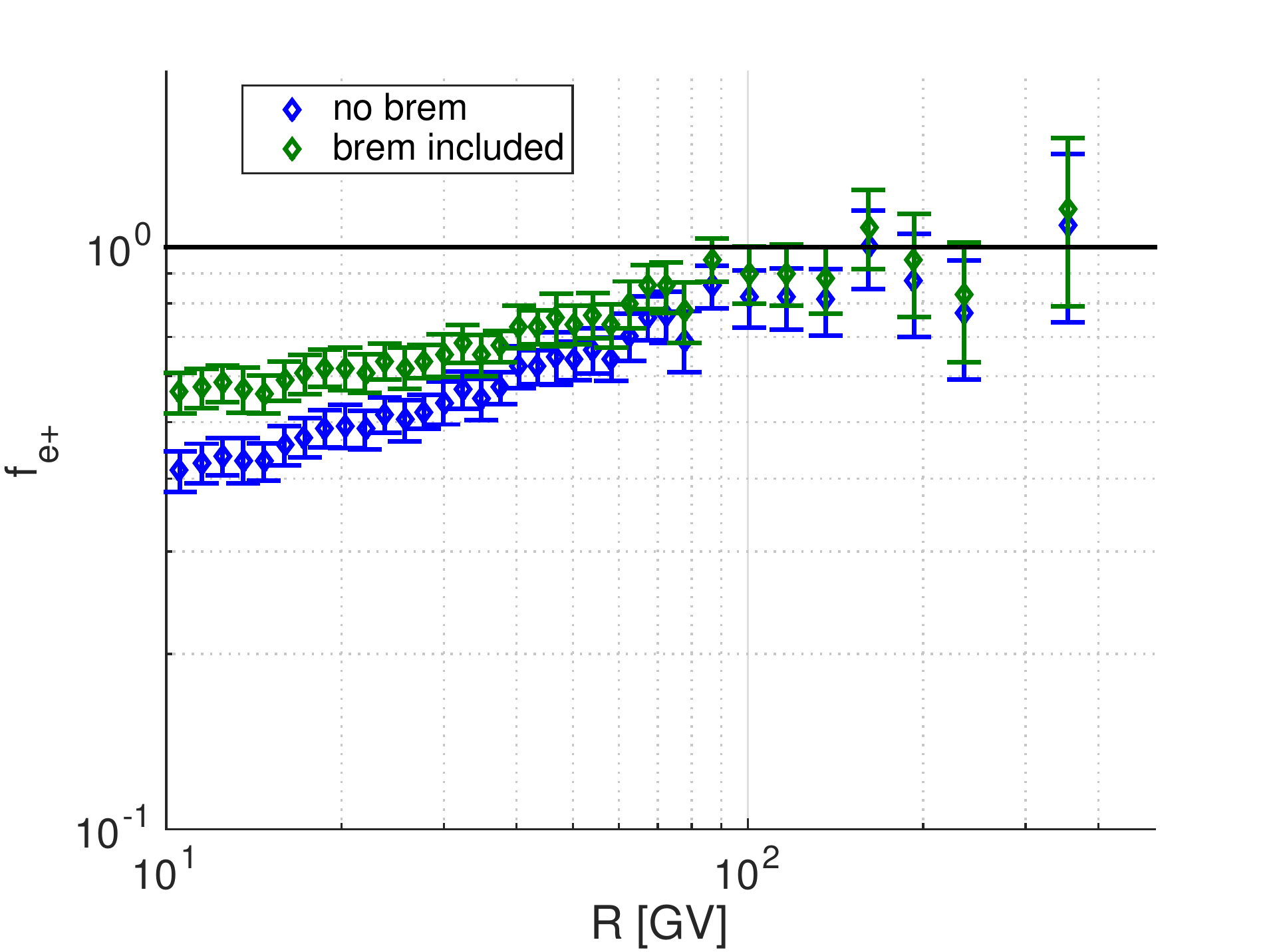}
\caption{$f_{\ep}$ extracted from Fig.~\ref{fig:pb2pos}. Green markers: bremsstrahlung losses estimated from $\X$ and subtracted from $f_{\ep}$. Blue markers same as in Fig.~\ref{fig:fep}.}
\label{fig:fepBrem}
\end{center}
\end{figure}

Using the similarity to fragmentation losses of nuclei, the $\ep$ loss suppression factor due to brem is
\be\label{eq:bremf} f_{\ep,\rm brem}&\approx&\frac{1}{1+\X\,\Gamma_{\rm brem}}.\ee
In Fig.~\ref{fig:fepBrem} we repeat the calculation of $f_\ep$, modding out (in green) the brem contribution using Eqs.~(\ref{eq:Gbrem}-\ref{eq:bremf}) with $C=2$. Bremsstrahlung modifies $f_\ep$ by $\sim40\%$ at $\R=10$~GV but becomes negligible at $\R\sim100$~GV.

%
We can also estimate the average ISM density traversed by CRs. Using Eq.~(\ref{eq:X}) together with Eqs.~(\ref{eq:tetc200}) and~(\ref{eq:tetc10}), we find
\be\bar n_{ISM}\left(\R=300\,{\rm GV}\right)\label{eq:n200}&\gtrsim&1\,\left(\frac{\bar U_T}{{\rm 1~eVcm^{-3}}}\right)\,{\rm cm^{-3}},\;\;\;\\
\bar n_{ISM}\left(\R=10\,{\rm GV}\right)\label{eq:n10}&\lesssim&0.15\,\left(\frac{\bar U_T}{{\rm 1~eVcm^{-3}}}\right)\,{\rm cm^{-3}},\;\;\;
\ee
assuming an ISM composition of 90\%H+10\%He by number.

Eqs.~(\ref{eq:n200}) and~(\ref{eq:n10}) suggest that the confinement volume of CRs may be decreasing with increasing CR rigidity, to the extent that CRs at $\R\sim300$~GV spend much of their propagation time within the thin Galactic HI disc, with a scale height $h\simeq200\,{\rm pc}$, while CRs at $\R\sim10$~GV probe a larger halo~\cite{Katz:2009yd}\footnote{This can also be stated as saying that the higher rigidity CRs escape the confinement volume more easily, and fail to return from a scale height that can still trap lower rigidity particles.}. There are other possibilities, however. For example, if a significant fraction of the grammage $X_{\rm esc}$ is accumulated during a short time in relatively dense regions, e.g. near the CR source, then the halo could be larger. Significant energy dependence in $\bar U_T$ could further affect the interpretation. For example, $\bar U_T\propto \R^{-0.6}$, inspired by $X_{\rm esc}\sim \R^{-0.4}$, would allow for a rigidity independent $\bar n_{ISM}$. 

Finally, it is also possible that the CR distribution is not in steady state. In this case, the PAMELA and AMS02 $\ep$ may be teaching us about, e.g., a recent nearby burst of star formation and supernova explosions. Key guidelines for such models, that can be derived from Figs.~\ref{fig:pb} and~\ref{fig:pb2pos}, are that:
\begin{enumerate}
\item B, $\pb$, and $\ep$ appear to be secondaries from the same spallation episode, and
\item the bulk of the spallation must not have occurred more than a few Myr in the past. 
\end{enumerate}
In the next section we review some ideas along these lines.

\subsection{Models for secondary $\ep$ and $\pb$}\label{ssec:secep}
In this section we briefly review CR propagation models~\cite{Katz:2009yd,Cowsik:2009ga,Burch:2010ye,Cowsik:2013woa,Blasi:2009hv,Blasi:2009bd,Mertsch:2009ph,Ahlers:2009ae,Kachelriess:2011qv,Cholis:2017qlb,Shaviv:2009bu,Kachelriess:2015oua,Thomas:2016fcp} where $\ep$ come from secondary production.  

In the nested leaky box model of~\cite{Cowsik:2009ga,Burch:2010ye,Cowsik:2013woa}, the secondary reacceleration model of~\cite{Blasi:2009hv,Blasi:2009bd,Mertsch:2009ph,Ahlers:2009ae,Kachelriess:2011qv,Cholis:2017qlb}, and the recent supernova model of~\cite{Kachelriess:2015oua,Thomas:2016fcp},  a common theme is that the spectrum of primary CRs in the secondary production sites is different than the locally measured spectrum. As a result, the application of Eq.~(\ref{eq:sec}) for relating the B/C garmmage $\X$ to secondary $\pb$ and $\ep$ becomes inaccurate (see the discussion around and below Eq.~(\ref{eq:pbcs})). This means that the success of Eq.~(\ref{eq:pbfromB}), seen in Fig.~\ref{fig:pb} to reproduce the $\pb/p$ ratio at $\R\lesssim100$~GV based on locally measured proton and nuclei spectra, should be somewhat accidental in these models. At $\R\gtrsim100$~GV, Fig.~\ref{fig:pb} is consistent with the possibility of a contribution to the spallation from a harder proton spectrum, although the systematic cross section and primary flux uncertainties prevent a sharp conclusion. 

We emphasize that the asymptotic $\ep/\pb$ ratio (that is the ratio when $\ep$ losses are not important) is insensitive to primary spectrum details, as can be seen in Fig.~\ref{fig:pb2pos}. 

The spiral arm model of~\cite{Shaviv:2009bu} is an example to a set-up in which $e^-$ and $\ep$ come from different regions in the Galaxy and thus their propagation energy losses are different. 

Ref.~\cite{Katz:2009yd} pointed out that a rigidity-dependent CR propagation volume could break the proportionality between $\te$ and $\X$, in accordance with the discussion of Sec.~\ref{sec:tetc}. Considering the diffusion model of Sec.~\ref{sec:tetc} (with more details in App.~\ref{app:demo}), for example, the idea is to let the boundary condition vary as $L=L(\R)$. Using Eqs.~(\ref{eq:tescdiff}) and~(\ref{eq:Xediff}) with $K\sim\R^\delta,\,L\sim\R^{-\delta_L}$, we  have $\te^{\rm diff}(\R)\sim\X^{\rm diff}(\R)\times\R^{-\delta_L}$. Fixing $\Xe\sim\R^{-0.4}$ to comply with B/C and $\pb/p$ data, setting $\delta_L\sim0.4,\,\delta\sim0$ the model can accommodate an $\ep$ loss suppression factor $f_{\ep}$ that is flat as function of $\R$ for $\tc\sim\R^{-0.8}$ (implying some rigidity dependence of $\bar U_T$ due to, e.g., Klein-Nishina effects and bremsstrahlung; see  Sec.~\ref{ssec:secepprop}). 
In the diffusion model, rigidity-dependent $L$ corresponds to non-separable rigidity and spatial dependence of the diffusion coefficient (the free escape boundary $L$ in these models mimics a region where the diffusivity diverges, $K\to\infty$)\footnote{Repeating footnote~15, the inhomogeneous diffusion model of~\cite{Kappl:2016qug} is not a good example for the set-up under discussion because it assumed a separable $\R$ and $z$ dependence of the diffusion coefficient. Thus $\te\propto\X$ in that model, just as for the simple homogeneous model with free escape boundary.}. Considering the picture of resonant pitch-angle scattering of CR on magnetic field irregularities~\cite{Blandford:1987pw}, non-separable rigidity and spatial dependence of $K$ means that the spectrum of magnetic field turbulence varies in the propagation region. $L$ decreasing with increasing $\R$ would occur if large scale turbulence decays faster than small scale turbulence at increasing distance from the Galactic disc. 
A rigidity-dependent confinement volume could also be realised in other settings~\cite{Blum:2010nx}. More quantitative analysis is required to test the idea further. 

Further discussion of~\cite{Cowsik:2009ga,Burch:2010ye,Cowsik:2013woa,Shaviv:2009bu,Blasi:2009hv,Blasi:2009bd,Mertsch:2009ph,Ahlers:2009ae,Kachelriess:2011qv,Cholis:2017qlb} and of the idea of a rigidity-dependent CR confinement volume can be found in~\cite{Katz:2009yd}. 
In the rest of this section we highlight the more recent model of~\cite{Kachelriess:2015oua,Thomas:2016fcp}.

Ref.~\cite{Kachelriess:2015oua,Thomas:2016fcp} suggested that a  supernova (SN) explosion occurring about 2~Myr ago at a distance of $\sim$100~pc from the solar system and injecting of the order of $10^{50}$~erg in CR protons could affect the spectra of primary and secondary CRs. The rate of SNe in the Milky Way is in the ballpark of 3 per century~\cite{Adams:2013ana}. Divided by the gas disc area $A_{\rm dsc}\sim10^3$~kpc$^2$, this gives $\sim3$ SNe per (300~pc)$^2$ per Myr, consistent with the set-up in~\cite{Kachelriess:2015oua}. The detection of CR $^{60}$Fe~\cite{Fry:2016hmb,Binns:2016wks} is also consistent with a recent nearby SN. 

Because $\pb$ and $\ep$ in~\cite{Kachelriess:2015oua} are produced as secondaries and their propagation time -- given roughly by the time since the SN -- is of order a Myr such that $\ep$ energy losses are not important below a few hundred GeV, the coincidence of the $\ep/\pb$ flux ratio with the secondary production rate ratio $Q_{\ep}/Q_{\pb}$ could be naturally addressed, as long as the recent SN contribution dominates for both species. 

The SN-originated proton flux in~\cite{Kachelriess:2015oua} dominates the local proton flux by a factor of $\gtrsim3$ at $\R\lesssim1$~TV, with the remaining flux assumed to come from an long time average of CR production in multiple earlier CR injection events. At the same time, the column density associated to the SN protons is $X\sim0.3$~g/cm$^2$, about 30\% of $\X$ derived from local B/C at $\R=1$~TV (see Fig.~\ref{fig:X}). As a result of this tuning (large SN proton flux with small associated grammage, vs. small average proton flux with large grammage), the SN-related contribution to the $\pb$ flux is about $\sim50$\% of the total at $\R\sim100$~GV. 

In the model of Ref.~\cite{Kachelriess:2015oua}, the CR distribution exhibits large deviations from steady-state with local sources producing order of magnitude deviations on sub-kpc and sub-Myr distance and time scales, as compared to the large volume or long time average. This picture can be tested with gamma ray data, by studying the gamma ray emissivity of individual molecular clouds and comparing to average diffuse emission data. Studies along these lines are ongoing~\cite{Ackermann:2012ik,Dermer:2013cfu,Strong:2015yva,Mizuno:2016rds,Neronov:2017lqd}.

\subsection{On dark matter and pulsar models for primary $\ep$ or $\pb$}\label{ssec:primep}

Many studies in the literature ascribed the CR $\ep$ or $\pb$ flux to a primary source, the most common examples being pulsars (see, e.g.~\cite{Profumo:2008ms,Hooper:2008kg,Malyshev:2009tw,Blasi:2010de,DiMauro:2014iia,Lee:2011zzs} and references to and therein) and dark matter annihilation (see, e.g.~\cite{Gaskins:2016cha,Meade:2009iu,Cholis:2013psa,Hooper:2009gm} and references to and therein).
In these works the primary source is assigned free parameters to describe the $\ep$ or $\pb$ spectrum and injection rate into the ISM. Then, using some CR diffusion model to simulate propagation effects, the model parameters are adjusted to fit the observed antimatter flux. 

%


While we do not think that current $\pb$ or $\ep$ data motivate the introduction of primary sources beyond the secondary flux, the {\it possibility} of dark matter or pulsar contributions is nevertheless interesting enough to merit some consideration.

The cosmological evidence for dark matter (DM) is compelling, and it is natural to imagine that DM is also responsible for the flattening of galactic rotation curves and other astrophysical gravitational anomalies. If DM is composed of massive particles, then pair annihilation of these particles in the Galactic halo could indeed produce high energy CR antimatter (examples predating the PAMELA data include, e.g.~\cite{Bottino:1998tw,Feng:2000zu,Donato:2003xg}). 
It is interesting to compare the CR antimatter source characterising a simple, generic DM model, to the irreducible secondary background. 
Focusing on $\pb$ for concreteness, the injection rate density from DM particle-anti-particle pair annihilation can be written as
\be\label{eq:DMpb} q_{\pb,\rm DM}&=&\frac{n_\chi^2(\vec r)\,\overline{\sigma v}}{4}\,\frac{dN_{\pb}}{dE}\\
&\approx&7\times10^{-34}\left(\frac{m_\chi}{100~\rm GeV}\right)^{-3}\frac{dN_{\pb}}{dx}\left(\frac{\overline{\sigma v}}{3\times10^{-26}~\rm cm^3/s}\right)\left(\frac{\rho_\chi(\vec r)}{0.3~\rm GeV/cm^3}\right)^2~{\rm cm^{-3}s^{-1}GeV^{-1}}.\no\ee
Here $n_\chi$ and $\rho_\chi=m_\chi n_\chi$ are the DM number and mass densities (see e.g.~\cite{Bahcall:1983bu,Navarro:1995iw,Merritt:2005xc,Sellwood:2008bd,RomanoDiaz:2008wz,Bovy:2012tw}) related by the DM mass $m_\chi$, $\overline{\sigma v}$ is the velocity-weighted mean annihilation cross section~\cite{Jungman:1995df}, and $dN_{\pb}/dx$ with $x=E/m_{\chi}$ is the (dimensionless) differential energy spectrum of $\pb$ produced per annihilation event. 

\begin{figure}[t]
\begin{center}
\includegraphics[scale=0.375]{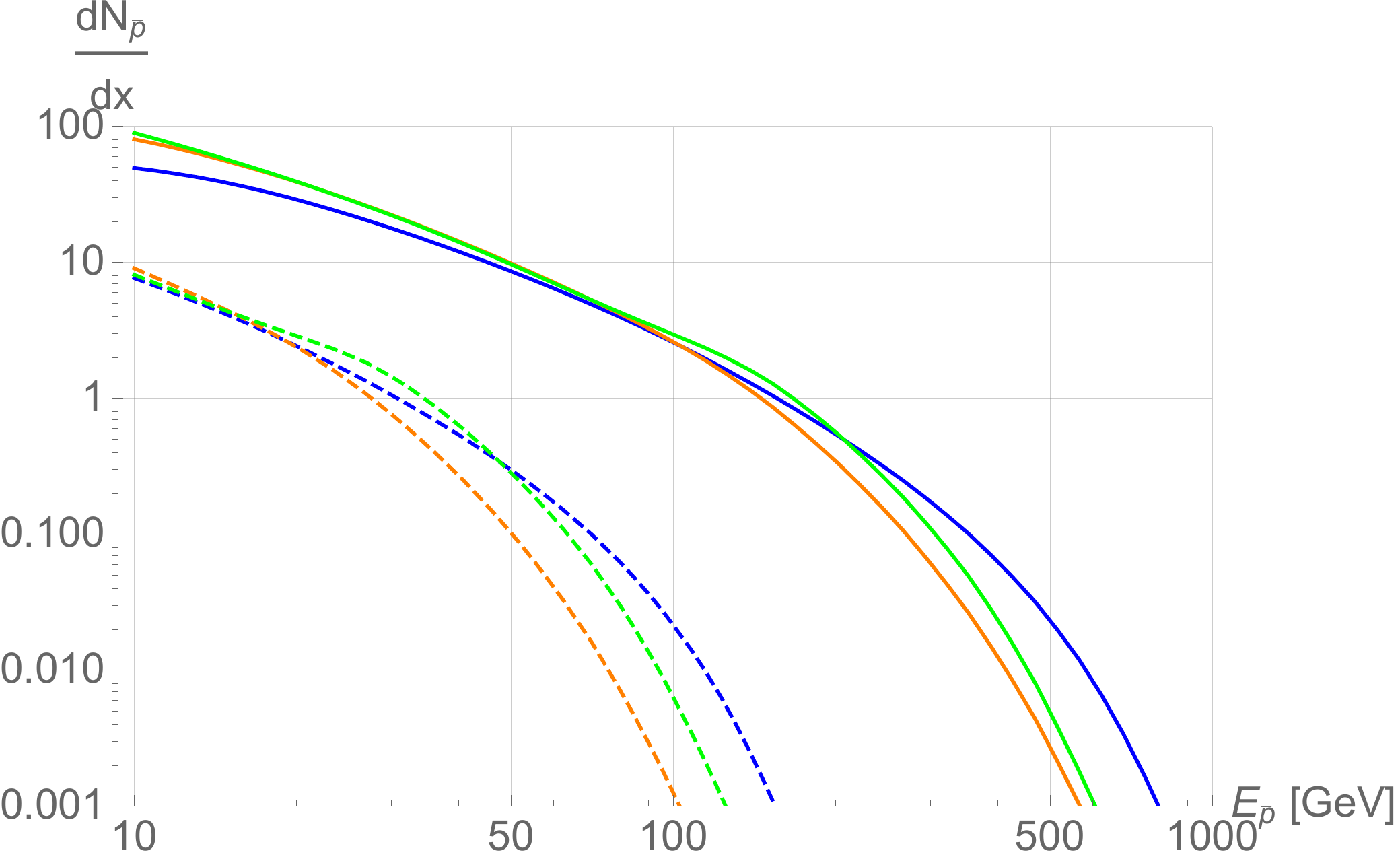}
\includegraphics[scale=0.375]{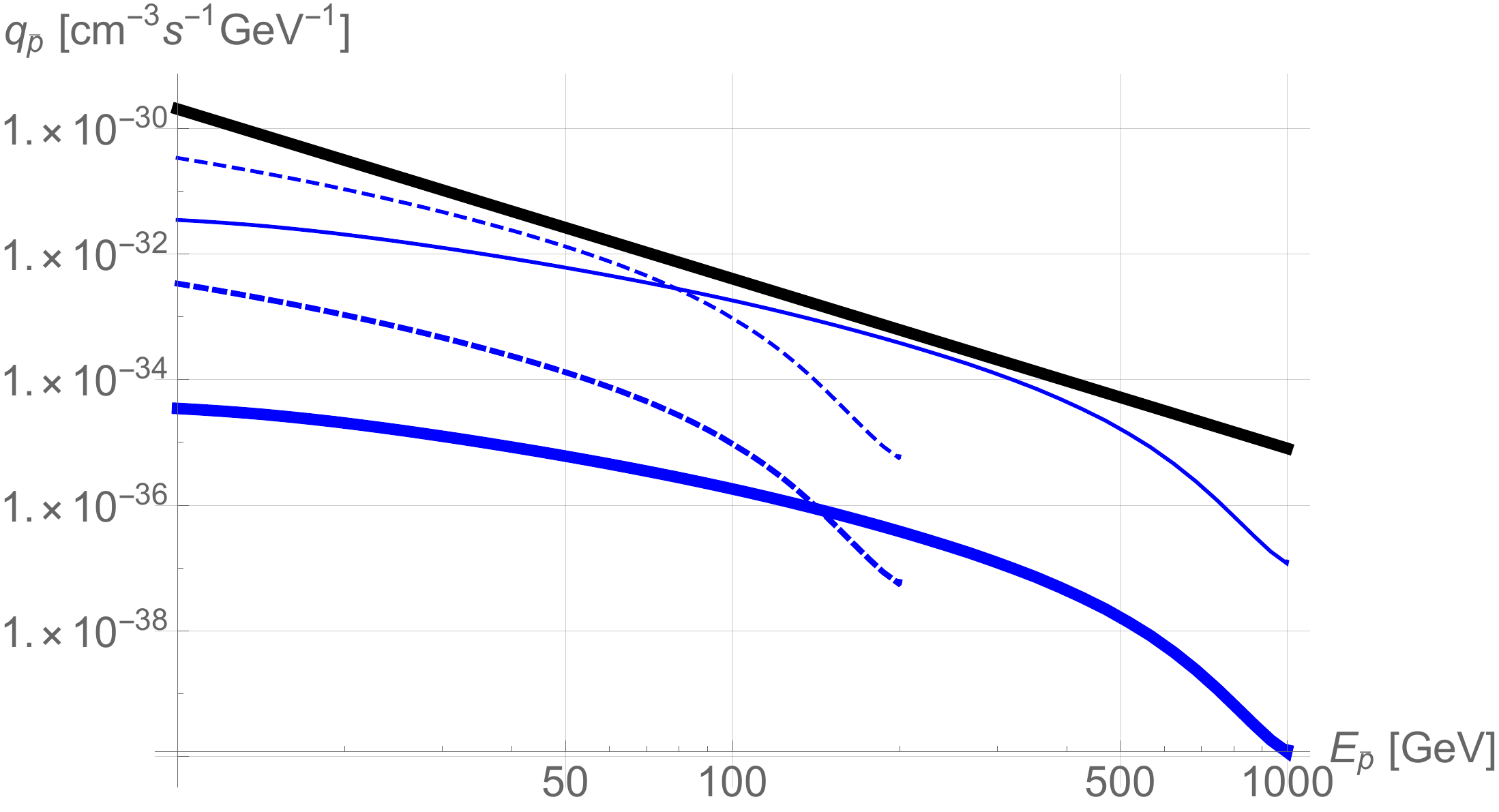}
\caption{{\bf Left:} 
Dimensionless spectra $dN_{\pb}/dx$ for $\pb$ produced in DM annihilation, calculated for $\chi\bar\chi\to W^+_LW^-_L$ (blue), $\chi\bar\chi\to hh$ (orange), $\chi\bar\chi\to b\bar b$ (green). Solid (dashed) lines show the result for $m_\chi=1$~TeV ($m_\chi=200$~GeV). 
{\bf Right:} Comparison of the secondary source rate density (black) computed with the approximate Eq.~(\ref{eq:qpbapp}), to DM annihilation source rate density. The DM annihilation spectra are for the $\chi\bar\chi\to W_L^+W_L^-$ channel and for DM mass of 1~TeV (solid) and 200~GeV (dashed). The thick, lower, DM lines show the result of Eq.~(\ref{eq:DMpb}), corresponding to a generic thermal relic DM model. The thin DM lines correspond to a multiplication of the output of Eq.~(\ref{eq:DMpb}) by a factor of 1000 for the 1~TeV case and 100 for the 200~GeV case.}
\label{fig:DMpb}
\end{center}
\end{figure}

The dimensionless spectrum $dN_{\pb}/dx$ is typically suppressed close to the kinematical endpoint at $x=1$, i.e. $dN_{\pb}/dx(x\approx1)\ll1$. To illustrate this we plot, in the left panel of Fig.~\ref{fig:DMpb}, $dN_{\pb}/dx$ as obtained for three sample annihilation channels: $\chi\bar\chi\to W^+_LW^-_L$ (blue), $\chi\bar\chi\to hh$ (orange), $\chi\bar\chi\to b\bar b$ (green). Solid (dashed) lines show the result for $m_\chi=1$~TeV ($m_\chi=200$~GeV).

The DM injection rate density can be compared to the secondary injection rate density,
\be\label{eq:qpbapp} q_{\pb,\rm sec}&=&\rho_{ISM}c\,Q_{\pb}\\
&\approx&4\times10^{-33}\left(\frac{E}{100~\rm GeV}\right)^{-2.7}~{\rm cm^{-3}s^{-1}GeV^{-1}},\no\ee
where $Q_{\pb}$ is taken from Eqs.~(\ref{eq:Q}) and~(\ref{eq:pbcs}) and $\rho_{ISM}\approx2\times10^{-24}$~g/cm$^3$ is a typical ISM mass density in the Galactic gas disc.

In the right panel of Fig.~\ref{fig:DMpb} we compare the DM injection rate density and the secondary injection rate density, taking the $\chi\bar\chi\to W_L^+W_L^-$ channel to represent the DM scenario.
The thick, lower, DM lines show the result of Eq.~(\ref{eq:DMpb}), corresponding to a generic thermal relic DM model. The thin DM lines correspond to a multiplication of the output of Eq.~(\ref{eq:DMpb}) by a factor of 1000 for the 1~TeV case and 100 for the 200~GeV case. The irreducible secondary source is shown in black.

We conclude that a generic, thermal relic weakly interacting DM model predicts an $\pb$ production rate density that is 2-3 orders of magnitude below the irreducible astrophysical secondary source as it occurs in a typical region in the Galactic gas disc. The picture for $\ep$ from DM annihilation is similar. On top of the source  estimate, the CR flux resulting from the DM source enjoys a model-dependent enhancement factor compared with the secondary flux, if the DM halo extends over a large volume above and below the thin Galactic gas disc where the secondary spallation occurs. This enhancement factor could range from a factor of few to a factor of $\sim$100, given roughly by the ratio of the CR propagation volume to the volume of the gas disc, with some dependence on the unknown details of the DM density profile (see App.~B in Ref.~\cite{Agashe:2009ja}). Even with this model-dependent volume enhancement factor, some enhancement mechanisms are often required to boost the DM annihilation cross section in a typical thermal relic DM model such that it could compete with the secondary background for CR energy above a few tens of~GeV; examples include, e.g.~\cite{Bovy:2009zs,Kuhlen:2009is,Robertson:2009bh}. The required large DM annihilation cross sections are constrained by cosmological data~\cite{Feng:2009hw,Madhavacheril:2013cna,Chan:2015gia,Ade:2015xua}, so that model building gymnastics is required to attribute observable high energy $\ep$ or $\pb$ flux to DM. \\

Finally, consider the idea of pulsars as the source of $\ep$. 
Pulsars prevail the Galaxy~\cite{TheFermi-LAT:2013ssa}; are likely producers of $e^\pm$ pairs~\cite{Goldreich:1969sb,Gruzinov:2014hga,Philippov:2017ikm}; and have been suggested as possibly detectable sources of CR $\ep$ before the PAMELA era~\cite{PhysRevD.52.3265,1989ApJ...342..807B}. Thus, invoking pulsars as the origin of CR $\ep$ is sometimes considered an Occum's Razor choice~\cite{Profumo:2008ms}. 
\begin{figure}[t]
\begin{center}
\includegraphics[scale=0.4]{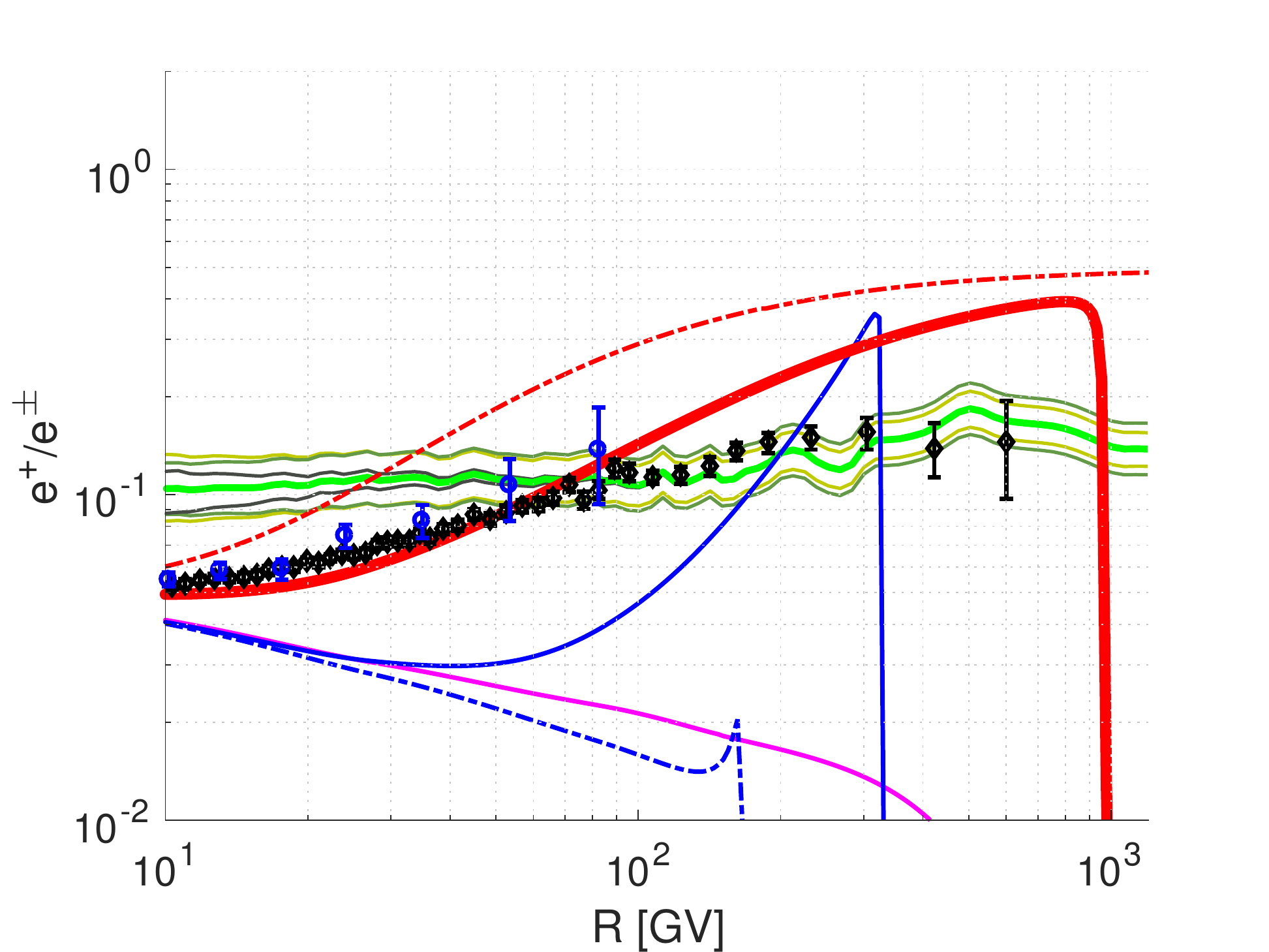}\quad
\includegraphics[scale=0.4]{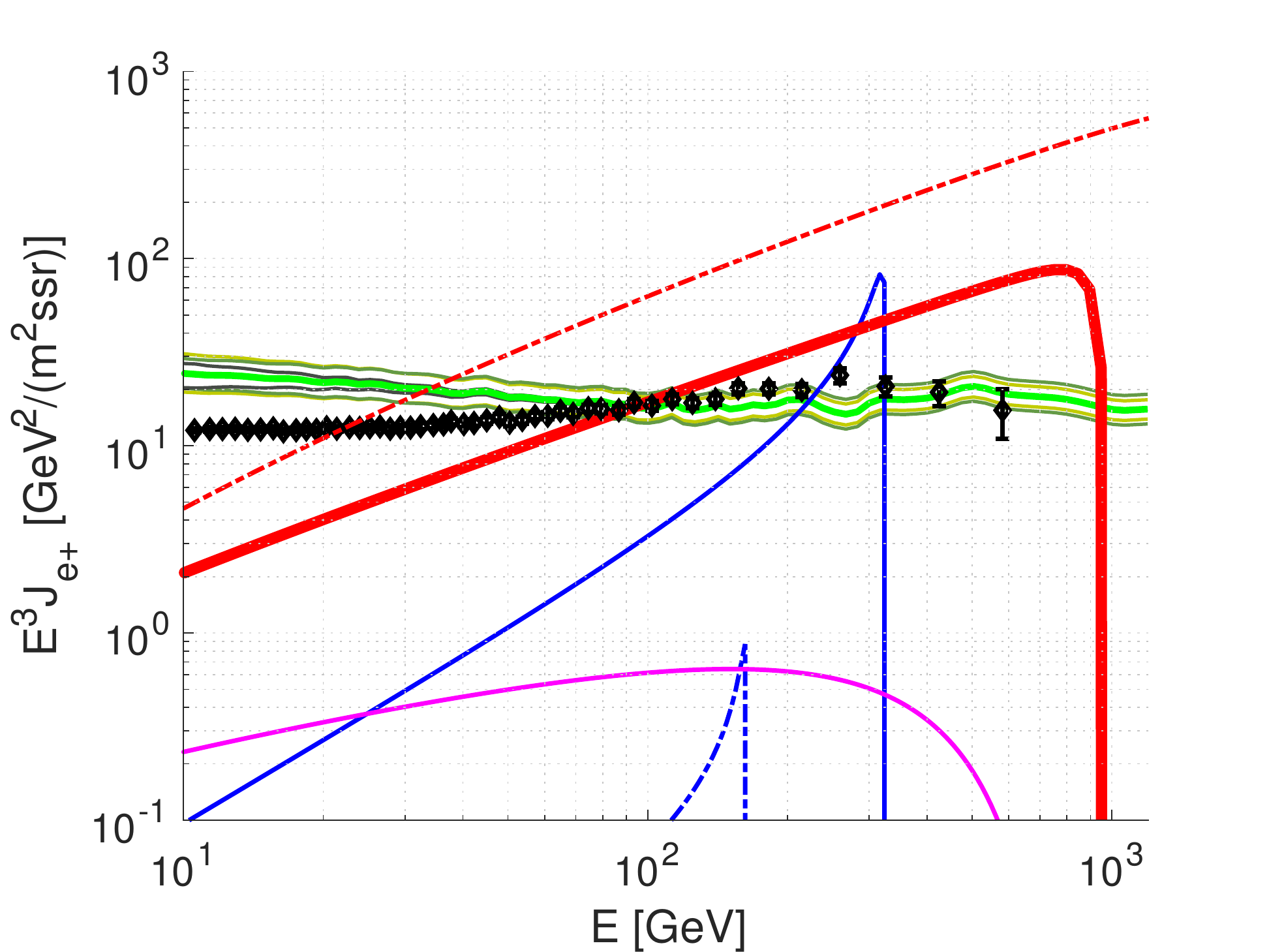}
\caption{Concerning a pulsar interpretation for CR $\ep$. {\bf Left:} $\ep/e^\pm$, {\bf Right:} $\ep$ flux. In both panels, the thick red line shows the output for a pulsar model that was fit in~\cite{Malyshev:2009tw} to match the then available PAMELA $\ep/e^\pm$ (blue markers on left)~\cite{Adriani:2008zr} and ATIC, HESS and FERMI $e^\pm$~\cite{Chang:2008aa,Aharonian:2008aa,Aharonian:2009ah,Abdo:2009zk} data. Additional lines show the output of the same model when free parameters are varied within part of the range deemed viable in~\cite{Malyshev:2009tw}; see text for details. The secondary $\ep$ upper limit derived from B/C data with no free parameters is shown in green.
}
\label{fig:pulsars}
\end{center}
\end{figure}

However, the production rate of $\ep$ by pulsars and the spectrum of the $e^\pm$ flux when it is finally released into the ISM are unknown. Therefore, pulsar models for CR $\ep$ take both the $\ep$ injection rate and spectrum as free, or at least poorly constrained parameters of the model, and fit these parameters to the $\ep$ data along with a set of propagation model parameters and along with additional free parameters to describe, as needed, the remaining primary $e^-$ flux on top of the pulsar contribution. 
With this phenomenological freedom in fitting the flux, the coincidence with the secondary upper limit seen in Figs.~\ref{fig:pb2pos}-\ref{fig:posfrac} is a surprise: a fine-tuned accident from the point of view of these models.

In Fig.~\ref{fig:pulsars} we demonstrate this point by comparing the prediction of a pulsar model from Ref.~\cite{Malyshev:2009tw} to recent AMS02 $\ep$ data. In both panels, the thick red line shows the output for a pulsar model that was fit in~\cite{Malyshev:2009tw} to match the then available PAMELA $\ep/e^\pm$~\cite{Adriani:2008zr} and ATIC, HESS and FERMI $e^\pm$~\cite{Chang:2008aa,Aharonian:2008aa,Aharonian:2009ah,Abdo:2009zk} data; see Fig.~2 in~\cite{Malyshev:2009tw}. 
To see what happens to this model when some (a-priori unknown) parameters are varied, we show in red dashed: the result if the assumed age of the pulsar is reduced by a factor of 3; magenta: assumed injection spectral index $n=1.99$ (vs. $n=1.6$ in Ref.~\cite{Malyshev:2009tw}'s fit), $e^\pm$ injection power reduced by factor of 3, and cut-off energy reduced by factor 20; blue smooth: $n=1$, injection power up by factor 5, age up by factor 3; blue dashed: $n=1$, injection power down by factor 5, age up by factor 6.

\section{Composite anti-nuclei: $\ad$ and $\ah$}\label{sec:adah}
Composite CR anti-deuterium ($\ad$) and anti-helium ($\ah$) have long been suggested as probes of dark matter~\cite{Donato:1999gy, Baer:2005tw, Donato:2008yx, Brauninger:2009pe, Kadastik:2009ts, Cui:2010ud, Dal:2012my, Ibarra:2012cc, Fornengo:2013osa, Carlson:2014ssa, Aramaki:2015pii}, as their secondary astrophysical production was thought to be negligible~\cite{Chardonnet:1997dv,Duperray:2005si,Ibarra:2013qt,Cirelli:2014qia,Herms:2016vop}. These references, and references to and within them, cover extensively the exciting possibility that dark matter annihilation or even primordial black hole evaporation could in principle produce a detectable flux of $\ad$ and/or $\ah$ in current and upcoming experiments such as GAPS~\cite{Aramaki:2015laa}, BESS~\cite{Abe:2011nx,2012PhRvL.108m1301A}, and AMS02~\cite{Giovacchini:2007dwa,kounineHebar}. Therefore, in the current review we do not enter further discussion of hypothetical exotic sources.

However, exotic sources aside, how does one actually predict the irreducible secondary flux? 

Using our tools from Sec.~\ref{sec:pbar}, CR propagation is not a serious difficulty when it comes to stable, relativistic, secondary nuclei -- and antinuclei, like $\pb$, $\ad$ and $\ah$. The challenge for CR $\ad$ and $\ah$ is set instead by inadequate particle physics data. Astrophysical anti-nuclei are dominantly produced in pp collisions, for which relevant cross section data is scarce when it exists at all. This has led attempts to calculate the flux of $\ad$ and $\ah$ into various extrapolations, resulting with large and difficult to quantify systematic uncertainty. 

\begin{figure}[t]
\begin{center}
\includegraphics[scale=0.41]{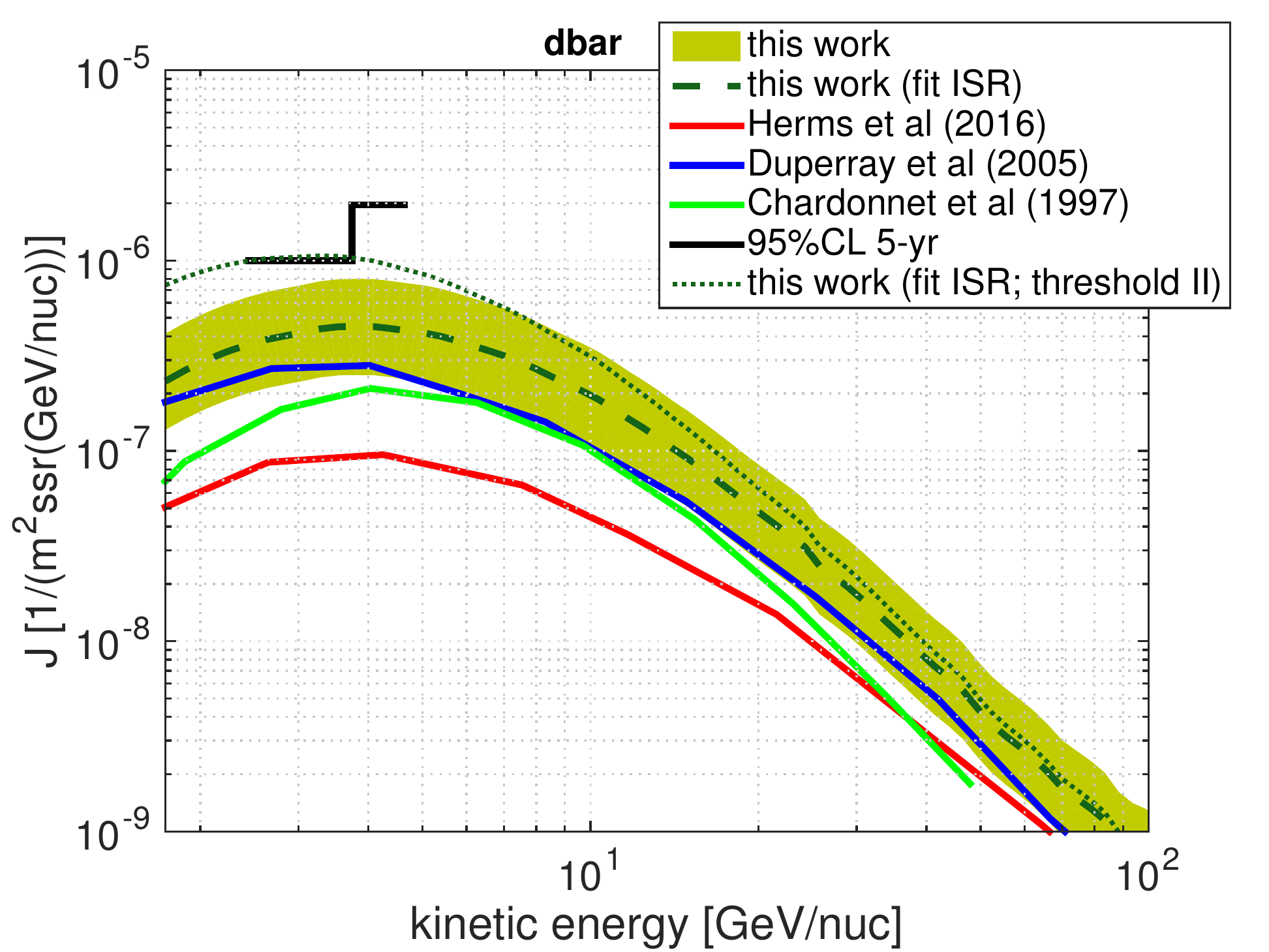}
\includegraphics[scale=0.41]{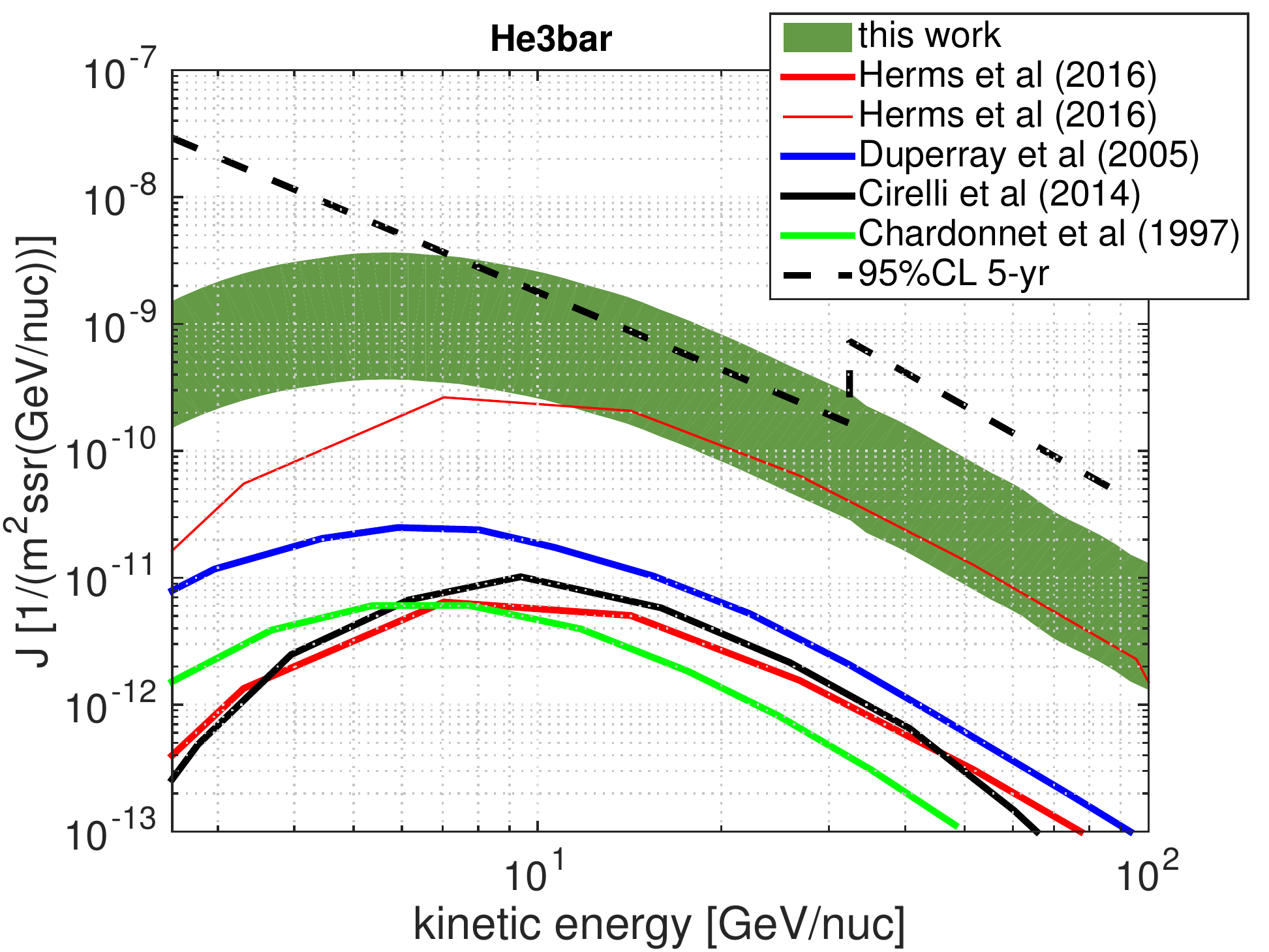}
\caption{{\bf Left:} estimates of the secondary CR $\ad$ flux. {\bf Right:} estimates of the secondary CR $\ah$ flux. Shaded bands denote the predictions of~\cite{Blum:2017qnn}, based on HBT data. AMS02 $\ad$ flux sensitivity (5-yr, 95\%CL) in the kinetic energy range 2.5-4.7~GeV/nuc, as estimated in~\cite{Aramaki:2015pii}, is shown in solid line. AMS02 $\ah$ flux sensitivity (5-yr, 95\%CL), derived from the $\ah/$He estimate of~\cite{kounineHebar}, is shown in dashed line. For more details, see~\cite{Blum:2017qnn}.}
\label{fig:dbahlit}
\end{center}
\end{figure}

A compilation of predictions of the secondary $\ad$ and $\ah$ fluxes from the literature is shown in Fig.~\ref{fig:dbahlit} (left and right panels, respectively). To date no detection of either $\ad$ or $\ah$ was officially announced by any experiment, although news of possible detection were reported by AMS02 in 2016.
\\

A recent attempt at tackling the cross section problem in $pp\to\ad,\,\ah$ production was done in~\cite{Blum:2017qnn}, which used a new technique burrowed from heavy ion femtoscopy~\cite{Lisa:2005dd,Scheibl:1998tk}. The $\ah$ flux predicted in~\cite{Blum:2017qnn}, shown by green band in the right panel of Fig.~\ref{fig:dbahlit}, is 1-2 orders of magnitude higher than most earlier estimates~\cite{Chardonnet:1997dv,Duperray:2005si,Ibarra:2013qt,Cirelli:2014qia,Herms:2016vop}.

The secondary $\ah$ flux could reach the 5-yr 95\%CL upper limit estimated for AMS02 prior to its launch~\cite{kounineHebar}.

Perhaps more important than the actual flux prediction, Ref.~\cite{Blum:2017qnn} scrutinised previous calculations of secondary $\ad$ and $\ah$ and highlighted extrapolations and possible sources of systematic uncertainties. 
In the rest of this section we outline this discussion. We show that LHC experiments are expected to shed light on these issues in the near future.\\

%
A coalescence ansatz~\cite{Butler:1963pp,Schwarzschild:1963zz,Gutbrod:1988gt} is often invoked to relate the formation of composite nucleus product with mass number $A$ to the formation cross section of the nucleon constituents:
\be\label{eq:coal}E_A\frac{dN_{A}}{d^3p_{A}}&=&B_A\,R(x)\,\left(E_{p}\frac{dN_{p}}{d^3p_{p}}\right)^A,\ee
where $dN_i=d\sigma_i/\sigma$ is the differential yield, $\sigma$ is the total inelastic cross section, and the constituent momenta are taken at $p_p=p_A/A$.  
The phase space factor $R(x)$, with $x=\sqrt{s+A^2m_p^2-2\sqrt{s}\tilde E_A}$ and $\tilde E_A$ the centre of mass product nucleus energy, is needed in order to extend the coalescence analysis down to near-threshold collision energies\footnote{See~\cite{Chardonnet:1997dv,Duperray:2002pj,Duperray:2003tv,Blum:2017qnn} for discussions; we use the evaluation of~\cite{Blum:2017qnn}.}.

Eq.~(\ref{eq:coal}) is useful to the extent that the coalescence factor $B_A$ is only mildly varying with initial state centre of mass energy (CME) and final state transverse momentum $p_t$.
Provided this condition is met, $B_A$ can be extracted from accelerator data in a kinematical or CME regime that is far from those directly useful to the astrophysics (e.g. at LHC experiments at high CME), and applied to astrophysics in the relevant kinematical regime using Eq.~(\ref{eq:coal}) and the more common $\pb$ and $\bar n$ production cross sections, that are reasonably well measured in a wide kinematical range.

However, experimental information on $\ad$ and $\ah$ production is scarce and, in the most part, limited to AA or pA collisions. For pp collisions, the most relevant initial state for CR astrophysics, no quantitative data exists 
for $pp\to\ah$, and the data for $pp\to\ad$ is sparse.\\

Faced with this problem, previous estimates~\cite{Chardonnet:1997dv,Duperray:2005si,Ibarra:2013qt,Cirelli:2014qia,Herms:2016vop} of the secondary CR $\ad$ and $\ah$ flux made two key simplifying assumptions: 
\begin{enumerate}
\item Coalescence parameters used to fit $pp\to\ad$ data were translated directly to $pp\to\ah$. The coalescence factor $B_A$ was converted to a coalescence momentum $p_c$, via
\be\label{eq:pcB} 
\frac{A}{m_p^{A-1}}\left(\frac{4\pi}{3}p_c^3\right)^{A-1}&=&B_A.\ee
The value of $p_c$ found from $pp\to\ad$ accelerator data was then assumed to describe $pp\to\ah$.
\item The same coalescence momentum was sometimes assumed to describe $pA\to\ad$ and $pp\to\ad$.
\end{enumerate}
Theoretical and empirical evidence suggests that both assumptions may be incorrect. To see this, we make a brief excursion into the physics of coalescence. \\

The role of the factor $B_A$ is to capture the probability for A nucleons produced in a collision to merge into a composite nucleus. It is natural for the merger probability to scale as~\cite{Bond:1977fd,Mekjian:1977ei,Csernai:1986qf} 
\be\label{eq:vscale}B_A\propto V^{1-A},\ee
where $V$ is the characteristic volume of the hadronic emission region. 
A model of coalescence that realises the scaling of Eq.~(\ref{eq:vscale}) was presented in Ref.~\cite{Scheibl:1998tk}. 
A key observation in~\cite{Scheibl:1998tk} is that the same hadronic emission volume is probed by Hanbury Brown-Twiss (HBT) two-particle correlation measurements~\cite{Lisa:2005dd}. Both HBT data and nuclear yield measurements are available for  AA and pA systems, allowing a test of  Eq.~(\ref{eq:vscale}). 

\begin{figure}[t]
\begin{center}
\includegraphics[scale=0.52]{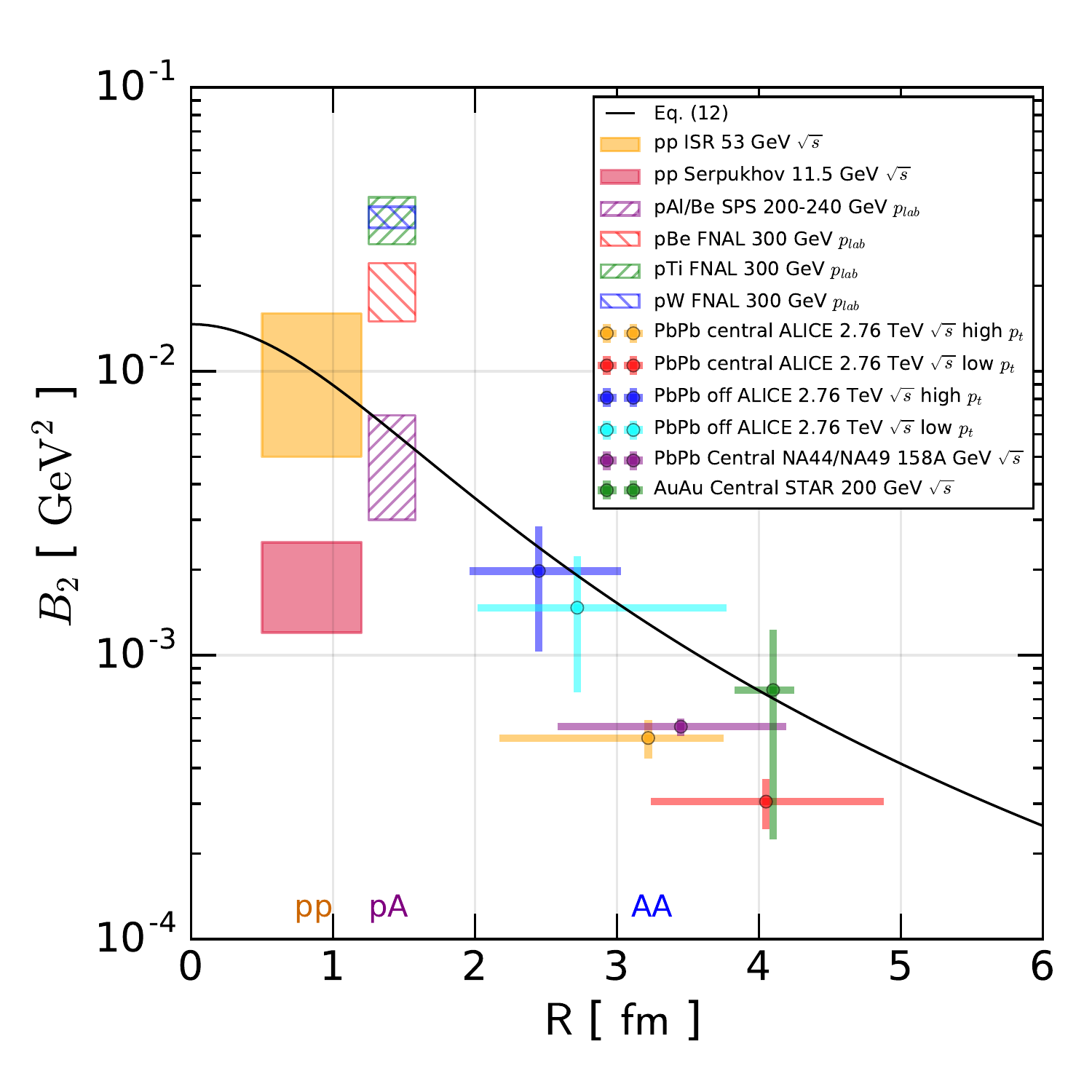}
\includegraphics[scale=0.52]{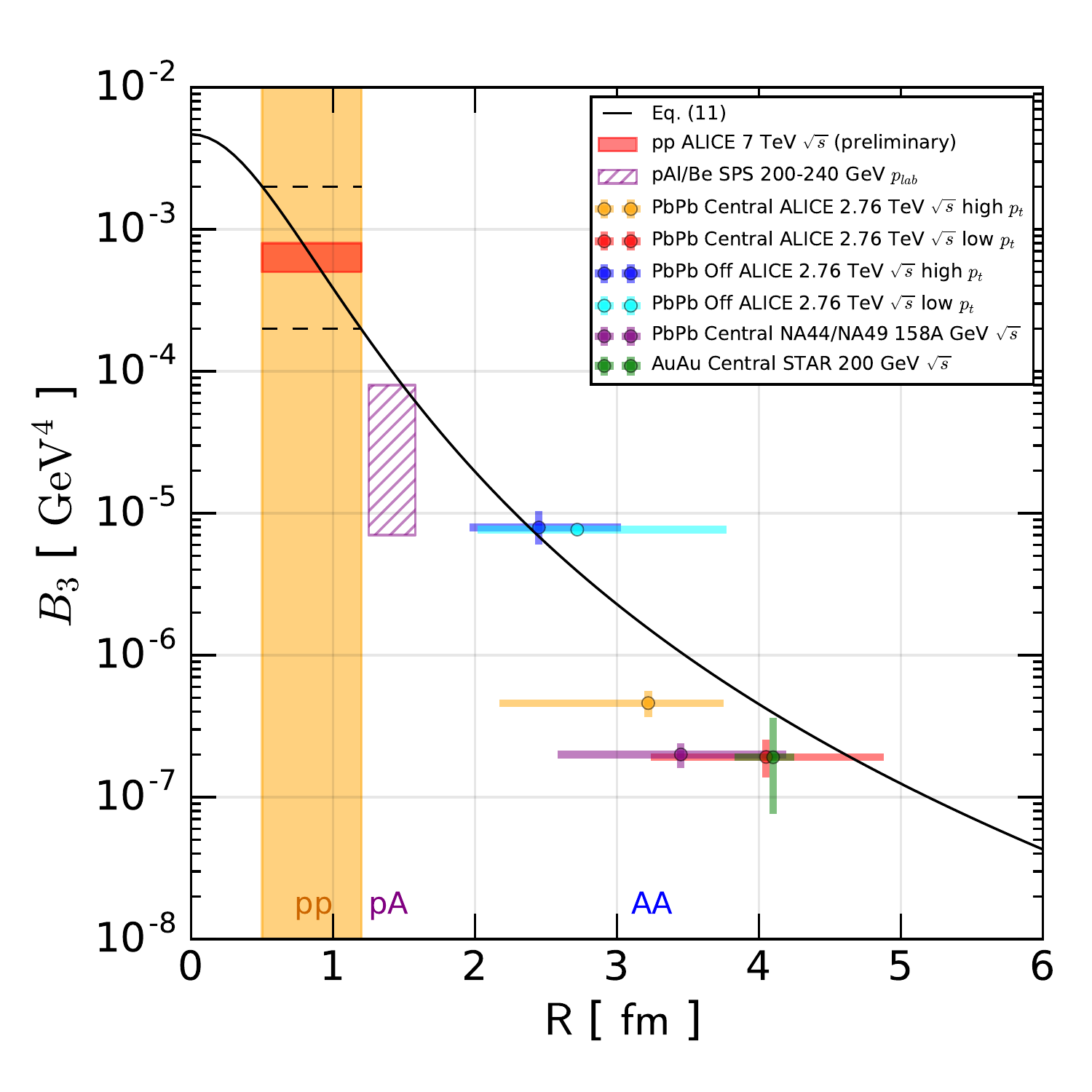}
\caption{Coalescence factor $B_2$ (Left) and $B_3$ (Right) vs. HBT radius. For more details, see~\cite{Blum:2017qnn}.}
\label{fig:Bformula}
\end{center}
\end{figure}

The coalescence factor in AA, pA, and pp collisions, presented w.r.t. HBT scale deduced for the same systems, is shown in Fig.~\ref{fig:Bformula}. The data analysis entering into making the plot is summarised in App.~A of~\cite{Blum:2017qnn}. 
The data is roughly consistent with Eq.~(\ref{eq:vscale}) as realised in~\cite{Scheibl:1998tk}, albeit with large uncertainty. 
\\

Importantly, Fig.~\ref{fig:Bformula} challenges the simplifying assumptions, utilised in one form or another in~\cite{Chardonnet:1997dv,Duperray:2005si,Ibarra:2013qt,Cirelli:2014qia,Herms:2016vop}, of using the same coalescence parameters for pp and pA collisions, or for $\ad$ and $\ah$ production. In particular, $\ah$ coalescence may well be more efficient than priviously estimated in these reference.\\

We can form a new estimate of $\ah$ production, and put the $\ad$ estimate in possibly useful context, by referring to Fig.~\ref{fig:Bformula}. 
HBT data for pp collisions~\cite{Uribe:1993tr,Malinina:2013fhp,Aamodt:2010jj} suggest $R$ in the range $0.5-1.2$~fm, indicated by letters in both panels of Fig.~\ref{fig:Bformula}. 
For $pp\to\ad$, direct measurements from the ISR~\cite{ALBROW1975189,1973PhLB...46..265A,Henning:1977mt} give
\be\label{eq:B2val} B_2^{(pp)}&=&(0.5-1.6)\times10^{-2}~{\rm GeV^2}.
\ee

For $pp\to\ah$ no direct experimental information is available to date. We extract a rough prediction of $B_3$, by taking the intersect of Eq.~(\ref{eq:vscale}), as realised in~\cite{Scheibl:1998tk}, with the two ends of the relevant range for $R$. This gives the following order of magnitude estimate:
\be\label{eq:B3val} B_3^{(pp)}&=&(2-20)\times10^{-4}~{\rm GeV^4}\;\;\,{\rm (HBT-\,based)},\ee
marked by the two horizontal dashed lines in the right panel of Fig.~\ref{fig:Bformula}. 

The $\ad$ and $\ah$ flux obtained with Eqs.~(\ref{eq:B2val}-\ref{eq:B3val}) and the grammage formalism of Eq.~(\ref{eq:sec}) are shown as shaded bands in the left and right panels of Fig.~\ref{fig:dbahlit}, respectively. \\

Very recently ALICE analysis\footnote{https://cds.cern.ch/record/2272148?ln=en} for pp collisions at $\sqrt{s}=13$~TeV was shown to reproduce $B_2^{(pp)}\approx1.6\times10^{-2}$~GeV$^2$, in agreement with Eq.~(\ref{eq:B2val}) and in support of the idea that the coalescence factor extracted at LHC energy can be usefully extrapolated to much lower CME. 

Results from ALICE~\cite{Sharma:2011ya} at $\sqrt{s}=7$~TeV pp collisions were analysed in~\cite{Blum:2017qnn}, and allow a preliminary test of Eq.~(\ref{eq:B3val}), indicating a consistent coalescence factor $B^{(pp)}_3\approx(5-8)\times10^{-4}$~GeV$^4$. A dedicated analysis by the ALICE collaboration is highly motivated and results are expected in the near future. LHCb~\cite{Alves:2008zz} is also conducting promising analyses of composite nuclei formation.

\section{Summary}\label{sec:sum}
We reviewed the theoretical interpretation of recent measurements of CR antimatter $\pb$ and $\ep$, along with some expected upcoming attempts to detect $\ad$ and $\ah$. While CR antimatter is a promising discovery tool for new physics or exotic astrophysical phenomena, an irreducible background arises from secondary production by primary CR collisions with interstellar matter. Understanding this irreducible background or constraining it from first principles is an interesting challenge. 

The state of the art cosmic ray measurements, currently dominated by results from the AMS-02 experiment~\cite{PhysRevLett.110.141102,Aguilar:2016kjl,AMS02C2O:2016}, lead to the following conclusions.
\begin{itemize}
\item  {\bf CR $\pb$ are most likely secondary}, coming from CR-gas collisions. This conclusion is based on comparison of the flux of secondary CR boron (B) and $\pb$. The $\pb$ and B fluxes are both consistent with secondary production due to primary CRs traversing the same amount of ISM column density. Variations to this simple picture, due e.g. to non-uniform CR elemental composition at the secondary production sites, are still possible but constrained by the data to the $\mathcal{O}(10\%)$ level. A key uncertainty in the analysis is due to the scarcity of high energy nuclear fragmentation cross section data, that is required for extracting the traversed column density and where all current astrophysical analyses are based on extrapolation from low energies.
\item {\bf CR $\ep$ data is consistent with the same secondary production mechanism responsible for $\pb$}. AMS02 measurements of the $\ep/\pb$ ratio reveal that this ratio is always below; comparable to; and saturates at high energy the ratio of the secondary production rates $Q_{\ep}/Q_{\pb}$, describing the branching fraction for $\ep$ and $\pb$ production in proton-ISM collisions. Given that $\pb$ are secondary or at least dominated by secondary production, it appears natural to conclude that $\ep$ are also secondary. The alternative hypothesis -- models for primary $\ep$ such as dark matter annihilation or pulsars -- must attribute the coincidence of $\ep/\pb$ with the secondary $Q_{\ep}/Q_{\pb}$ to a fine-tuned accident involving unrelated free model parameters. Secondary $\ep$ would imply that the CR propagation time $\te$ scales differently with CR rigidity than does the CR grammage $\X$, and is not much larger than 1~Myr at $\R\gtrsim100$~GV, suggesting that assumptions made in the context of common phenomenological diffusion models are incorrect. Existing radioactive nuclei data from HEAO3 are consistent with secondary $\ep$, and improved upcoming measurements by AMS02 would test the interpretation further.
\item {\bf The  flux of secondary high energy $\ah$ may be observable with a few years exposure of  AMS-02.} An indirect combined analysis of a large set of high energy accelerator data on pp, pA, and AA collisions suggests that the cross section for $pp\to\ah$ is larger than was estimated in most earlier CR literature, by a factor of 10-100. More direct measurements are expected soon from the LHC collaborations, notably ALICE and LHCb. If true, this analysis predicts the detection of secondary $\ah$ events in 5-10 years exposure of AMS02. Secondary $\ah$ events are constrained to high energy: no secondary event should be seen below $\sim1$~GeV/nuc. At the time of writing this review, preliminary reports by the AMS02 collaboration hint to a possible positive detection, but the analysis is not background-free and conclusive results are not yet available.
\end{itemize}
%

\mysections{Acknowledgments}
KB and RS are supported by the I-CORE program of the
Planning and Budgeting Committee and the Israel
Science Foundation (grant number 1937/12) and by grant 1507/16
from the Israel Science Foundation. KB is incumbent
of the Dewey David Stone and Harry Levine career
development chair.

\begin{appendix}
\section{Physical meaning of the CR grammage}\label{s:gram}
The cosmic ray grammage $\Xe$ (in units of column density) is a useful object for calculating the flux of relativistic, stable, secondary CR nuclei like B, sub-Fe, and $\bar{p}$. 
For these secondary nuclei, the following formula gives an empirical description of a large set of CR data~\cite{1988ApJ...324.1106B,Engelmann:1990zz,2001ApJ...547..264J,2003ApJ...599..582W}:
\be\label{eq:gra} n_i(\R)=Q_i(\R)\,\Xe(\R),\ee
where $Q_i$ is the local net source (production - losses) for species $i$ per unit column density traversed. Importantly, $\Xe$ is a universal function of rigidity and does not carry the species label $i$.

Eq.~(\ref{eq:gra}) has become known in the literature as the ``leaky box equation", and is often confused with the so-called leaky box model, which is a simplistic model of CR propagation in which the density of CRs and the density of ISM are  assumed to be distributed uniformly in the propagation region. However, Eq.~(\ref{eq:gra}) is more general and applies to a broad class of propagation models, of which the leaky box model is just one  example. Equally good examples of models that realize Eq.~(\ref{eq:gra}) are the commonly adopted steady-state homogeneous diffusion models where the CRs are assumed to diffuse in a large halo of rigidity-independent scale height $L$, enclosing a thinner Galactic gas disc of scale height $h\ll L$. As we show below, neither steady-state, nor diffusion (homogeneous or otherwise) nor special boundary conditions are required in order for Eq.~(\ref{eq:gra}) to apply. Needless to say, having verified Eq.~(\ref{eq:gra}) observationally, one cannot deduce that any of the above simplifying assumptions applies in Nature.\\

A sufficient condition for Eq.~(\ref{eq:gra}) to apply, is that the relative {\it composition} ({\it not} the density) of the CRs and the ISM be uniform in the time and place in the Galaxy at which secondary CRs are dominantly produced and from which they arrive to our time and place in the Galactic disc [denoted in what follows by $(\vec r_\odot,t_\odot)$]~\cite{Katz:2009yd}. Various propagation models in the literature satisfy this condition to good a approximation. 

To make things concrete we focus first on CR boron (B).
The net source for B, defined at an arbitrary point $(\vec r,t)$ in the Galaxy, is
\be\label{eq:QB} Q_B\x=\sum_{i={\rm C,N,O,...}}\left(\frac{\sigma_{i\to{\rm B}}}{m_{\rm ISM}}\right)n_i\x-\left(\frac{\sigma_{\rm B}}{m_{\rm ISM}}\right)n_B\x.\ee
Here $m_{\rm ISM}$ is the average mass per ISM particle (for 90\%H+10\%He we have $m_{\rm ISM}=1.3~m_p$, where $m_p=1.67\times10^{-24}$~g); $\sigma_{i\to{\rm B}}$ is the cross section for nucleus species $i$ to scatter on ISM particles and fragment into B; $\sigma_{\rm B}$ is the total fragmentation cross section of B. 
We have suppressed the energy dependence of the cross sections, taking the produced B in the reaction $i\to{\rm B}$ to inherit the rigidity of the parent CR $i$. Note that the energy dependence of the cross section could become important, particularly so when we will come to apply our results to secondary $\bar p$. There, we will need to generalise Eq.~(\ref{eq:QB}) as applied to $\bar p$ production. We will come back to this point later on; for now let us assume that, as in Eq.~(\ref{eq:QB}), we can describe the production of secondary CR at rigidity $\R$ by an effective cross section $\sigma$ multiplied by the primary CR particle density taken at the same rigidity $\R$.

\begin{figure}[!h]\begin{center}
\includegraphics[width=0.6\textwidth]{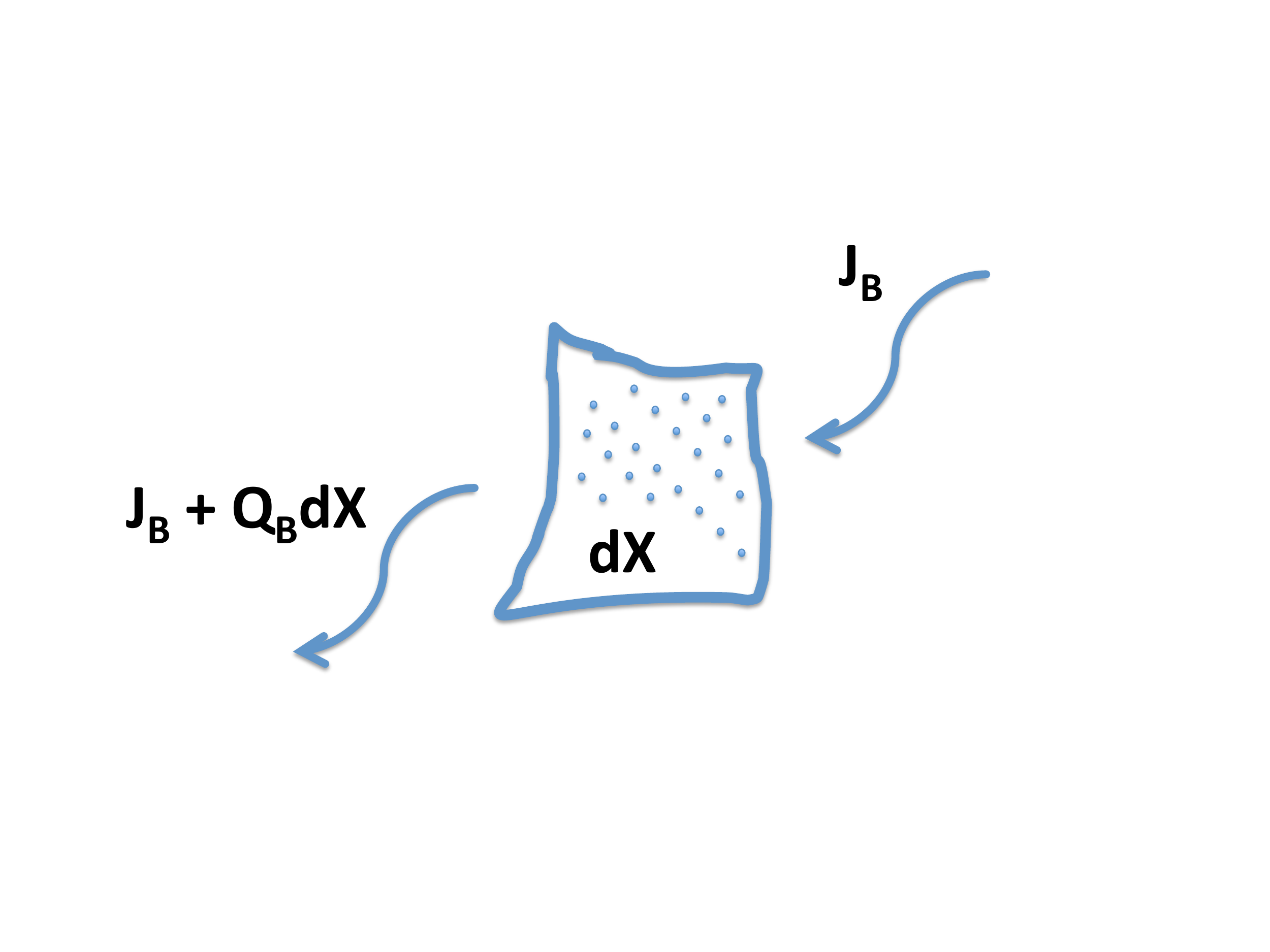}
\end{center}
\caption{Meaning of the net source $Q_B(\R,\vec r,t)$.}
\label{fig:Q}
\end{figure}

The {\it net} source, Eq.~(\ref{eq:QB}), accounts for fragmentation losses and not only for the production of the secondaries. This allows us to isolate the effect of propagation and put different secondary CR species on equal grounds even if their fragmentation cross sections are different, as is the case for B, sub-Fe, and $\bar p$. Upon traversing some volume element in the ISM, located around some point $(\vec r,t)$ with column density $dX$, the flux of B is augmented by a contribution $J_{\rm B}(\R,\vec r+d\vec r,t+dt)=J_{\rm B}(\R,\vec r,t)+Q_{\rm B}(\R,\vec r,t)\,dX$. This is illustrated in Fig.~\ref{fig:Q}.

To obtain the density of secondary CRs here and now, we sum the contributions $Q\,dX$ of volume elements throughout the Galaxy, weighted by the probability $P\left(\R,\{\vec r,t\},\{\vec r_\odot,t_\odot\}\right)$ of a CR with rigidity $\R$, produced at $(\vec r,t)$, to arrive to us at $(\vec r_\odot,t_\odot)$, {\it omitting fragmentation loss}. Because our use of the net source $Q$ instructs us to omit explicit fragmentation loss in $P$, and because propagation in a magnetic field depends only on the CR rigidity $\R$, we deduce that $P$ does not carry a species label and should apply equally well to B, sub-Fe, or $\bar p$. 
The locally measured density of B, $\nb(\R,\vec r_\odot,t_\odot)$, is therefore given by the expression:
\be\label{eq:int}\nb\xo=\int dt\int d^3r\,c\,\rism(\vec r,t)\,Q_{\rm B}\x\,P\left(\R,\{\vec r,t\},\{\vec r_\odot,t_\odot\}\right).\ee

Scaling out the locally measured net production rate, and highlighting the role of the local CR density  in the integrand [that we choose to show by the ratio of the density of carbon (C) at the point $(\vec r,t)$ compared to the locally measured value at $(\vec r_\odot,t_\odot)$], we have
\be\label{eq:X0} \frac{\nb\xo}{Q_{\rm B}\xo}&=&
\int dt\int d^3r\,c\,\rism(\vec r,t)\,\frac{\nc\x}{\nc\xo}\,P\left(\R,\{\vec r,t\},\{\vec r_\odot,t_\odot\}\right)\no\\
&\times&\frac{Q_{\rm B}\x/\nc\x}{Q_{\rm B}\xo/\nc\xo}.\ee

Let us examine the last factor in the integrand in Eq.~(\ref{eq:X0}):
\be\frac{Q_{\rm B}\x/\nc\x}{Q_{\rm B}\xo/\nc\xo}&=&\frac{\sum_{i={\rm C,N,O,...}}\left(\frac{\sigma_{i\to{\rm B}}}{m_{\rm ISM}}\right)\frac{n_i\x}{n_{\rm C}\x}-\left(\frac{\sigma_{\rm B}}{m_{\rm ISM}}\right)\frac{n_B\x}{n_{\rm C}\x}}{\sum_{i={\rm C,N,O,...}}\left(\frac{\sigma_{i\to{\rm B}}}{m_{\rm ISM}}\right)\frac{n_i\xo}{n_{\rm C}\xo}-\left(\frac{\sigma_{\rm B}}{m_{\rm ISM}}\right)\frac{n_B\xo}{n_{\rm C}\xo}}.\ee
Uniform CR composition means that the ratio $\frac{n_i\x}{n_{\rm C}\x}$, while it depends on the species labels $i$ and C, and on rigidity $\R$, does not depend on the point $(\vec r,t)$. We see that if the CR composition is uniform in the time and place in the Galaxy in which the integrand in Eq.~(\ref{eq:X0}) receives most of it's support, then we can set
\be\frac{Q_{\rm B}\x/\nc\x}{Q_{\rm B}\xo/\nc\xo}=1.\ee
In addition, we may write $n_C\x/n_C\xo= n_{\rm CR}\x/n_{\rm CR}\xo$, where by $n_{\rm CR}\x/n_{\rm CR}\xo$ we refer collectively to the density contrast of CRs of all species at point $(\vec r,t)$ compared to their density here at $(\vec r_\odot,t_\odot)$. Plugging these results into Eq.~(\ref{eq:X0}) we find, finally,
\be\nb\xo=Q_{\rm B}\xo\,\Xe(\R),\ee
which coincides with Eq.~(\ref{eq:gra}) and identifies $\Xe$:
\be\Xe(\R)\equiv\int dt\int d^3r\,c\,\rism(\vec r,t)\,\frac{n_{\rm CR}\x}{n_{\rm CR}\xo}\,P\left(\R,\{\vec r,t\},\{\vec r_\odot,t_\odot\}\right).\ee

We now get back to the issue of energy dependent fragmentation cross sections. For high energy ($E\gtrsim10$~GeV/nuc) nuclei, fragmentation is dominated by processes in which a parent nucleus loses an $\alpha$ particle or a few (dominantly one) nuclei. In this process, to a good approximation, the Lorentz factor $\Gamma$ of the parent nucleus is inherited by the leading daughter nucleus -- this is the straight-ahead approximation. Since, for the nuclei in question, $A\approx 2Z$ to about 10\% accuracy, the magnetic rigidity $\R$ is also inherited by the daughter nucleus to $\mathcal{O}(10\%)$ accuracy. Extending the analysis to CR $\pb$, where the main production cross section $pp\to\pb$ exhibits nontrivial kinematic dependence, is straightforward but introduces sensitivity to the proton spectrum in the secondary production regions. This was discussed in Sec.~\ref{sec:pb2pb2c}.\\

Early analyses~\cite{1988ApJ...324.1106B,Engelmann:1990zz,2001ApJ...547..264J,2003ApJ...599..582W} relying on HEAO3 data~\cite{Engelmann:1990zz} determined the value of  $X_{\text{esc}}$ up to $\R\sim300$~GV. Using recent AMS02 B/C data~\cite{AMS02ICRC} to extract the value of $X_{\text{esc}}$ up to $\R=1$~TV, Ref.~\cite{Blum:2013zsa} derived an approximate power law fit, 
\be\label{eq:X} X_{\rm esc}=8.7\left(\frac{\R}{10~{\rm GV}}\right)^{-0.4}\,{\rm g\,cm^{-2}},\ee
consistent at lower energy but slightly harder in slope than deduced from the earlier data~\cite{1988ApJ...324.1106B,Engelmann:1990zz,2001ApJ...547..264J,2003ApJ...599..582W}.
This result was later extended directly from data in our evaluation of Eq.~(\ref{eq:X}) as shown in the right panel of Fig.~\ref{fig:X}.

\section{Radiative energy loss of $\ep$ vs. CR propagation time: model examples}\label{app:demo}
The detailed interpretation of the form of the $\ep$ loss suppression factor $f_{e^+}$ is model-dependent. 
To demonstrate this point, we calculate $f_{e^+}$ for two propagation model examples. 

First, we consider a version of the leaky-box model (LBM). In this model we assume that the CR density is homogeneous inside some propagation volume $V_{CR}$, containing the MW gas disc and possibly extending some (possibly large) distance above it\footnote{Note that there is no reason to impose that the ISM or the sources of CR are also distributed uniformly in the propagation region. It would be enough that CRs bounce multiple times off the ``box" boundaries.}. We assume that CRs are trapped in $V_{CR}$ for a rigidity dependent time $t^{\rm LBM}_{\rm esc}(\R)\propto\R^{-\delta}$ before they escape, and that the average ISM mass density in the propagation region is $\langle\rho_{ISM}\rangle$. We assume that $V_{CR}$ itself is rigidity-independent, so $\langle\rho_{ISM}\rangle$ is also rigidity-independent. 

Second we consider a one-dimensional thin disc+halo homogeneous diffusion model, with diffusion coefficient $K(\R)\propto\R^\delta$ inside a CR propagation region extending to a distance $L$ above and below the gas disc of half-width $h\ll L$. We assume free escape boundary conditions at $z=\pm L$, and again assume that $L$ is rigidity-independent. 

Both propagation models satisfy Eq.~(\ref{eq:sec}). Therefore calibrating the free parameters in either model (that is, the function $t^{\rm LBM}_{\rm esc}(\R)$ and the quantity $\langle\rho_{ISM}\rangle$ for the LBM, or the function $K(\R)$ and the quantity $L$ in the diffusion model) to fit the measured B/C, results automatically in a consistent prediction for $\pb/p$ as seen in Fig.~\ref{fig:pb}.
We can calculate $\Xe$:
\be\label{eq:XeLBM} \Xe^{\rm LBM}(\R)&=&\langle \rho_{ISM}\rangle c\,t^{\rm LBM}_{\rm esc}(\R),\\
\label{eq:Xediff}\Xe^{\rm diff}(\R)&=&X_{\rm disc}\frac{L\,c}{2K(\R)}=\frac{X_{\rm disc}\,c}{L}\,t_{\rm esc}^{\rm diff}(\R).
\ee
Here $X_{\rm disc}=2h\rho_{ISM}$ is the grammage of the gas disc, and we define the CR propagation time for the diffusion model:
\be\label{eq:tescdiff} t_{\rm esc}^{\rm diff}(\R)&=&\frac{L^2}{2K(\R)}.\ee

Turning to the $\ep$ loss suppression factor, we find 
\be f_{e+}^{\rm LBM}(\R)&=&\tce\int_1^\infty dxx^{-\gamma_i}\exp\left[-\tce\frac{1-x^{\delta-1}}{1-\delta}\right]\no\\
&\longrightarrow&\frac{1}{\gamma_i-1}\tce,\\
f_{e+}^{\rm diff}(\R)&=&\sqrt{\tce}\sqrt{\frac{1-\delta}{\pi}}\int_1^\infty dx\frac{x^{-\gamma_i}}{\sqrt{1-x^{\delta-1}}}\sum_{m=-\infty}^{\infty}(-1)^m\exp\left[-\frac{1-\delta}{1-x^{\delta-1}}\tce m^2\right]\no\\
&\longrightarrow&\sqrt{\tce}C_{\rm diff}(\gamma_i,\delta),\;\;\;C_{\rm diff}(2.7,0.4)\approx0.8,
\ee
where we assumed that $t_{\rm cool}\propto\R^{-1}$, and that the $e^+$ are produced by a power-law population of protons with spectral index $\gamma_i$.  
The asymptotic behaviour at strong loss (obtained for $t_{\rm cool}\ll t_{\rm esc}$~\cite{Katz:2009yd}) is highlighted in the second line for each model.

While the diffusion model and the LBM differ in the form of $f_{e^+}$ they predict,  both models share a common feature: because the LBM  volume $V_{CR}$ was taken to be rigidity-independent, so is the average ISM density $\langle\rho_{ISM}\rangle$ explored by the CRs and thus $\Xe^{\rm LBM}(\R)\propto\te^{\rm LBM}(\R)$. Similarly, because the diffusion halo boundary $L$ in the diffusion model was taken to be rigidity independent, also here $\Xe^{\rm diff}(\R)\propto\te^{\rm diff}(\R)$. In other words: in both of the models, the rigidity-dependent column density of ISM traversed by CRs, is proportional to the rigidity-dependent CR propagation time.

\end{appendix}

\bibliography{ref}
%
%
%

\end{document}